\documentclass[aps,prd,twocolumn,superscriptaddress,showpacs,showkeys]{revtex4-2}
\usepackage{amsmath,amssymb}
\usepackage{color,xcolor}
\usepackage{graphicx,mathrsfs}
\usepackage{tensor}

\newskip\humongous \humongous=0pt plus 1000pt minus 1000pt

\newif\ifdtup

\relax

\newcommand{\bea}{\begin{eqnarray}}
\newcommand{\eea}{\end{eqnarray}}
\newcommand{\nn}{\nonumber}
\newcommand{\pro}{\partial}

\newcommand{\pd}{\partial}
\newcommand{\mn}{{\mu\nu}}

\newcommand{\bA}{{\bar A}}

\newcommand{\hD}{{\hat D}}

\newcommand{\cD}{{\cal D}}

\newcommand{\e}{{\vec e}}

\newcommand{\del}{\delta}
\newcommand{\vlam}{{\vec \lambda}}

\newcommand{\lam}{{\lambda}}

\newcommand{\vg}{{\bf g}}
\newcommand{\hvg}{\hat {{\bf g}}}
\newcommand{\vG}{{\bf G}}

\newcommand{\tG}{{\widetilde{G}}}
\newcommand{\Gm}{{\Gamma}}

\newcommand{\vGm}{{\bf \Gamma}}
\newcommand{\hvGm}{\hat{{\bf \Gamma}}}
\newcommand{\tGm}{{\widetilde{\Gamma}}}
\newcommand{\tH}{{\widetilde{H}}}
\newcommand{\vI}{{\bf I}}
\newcommand{\vj}{{\bf {j}}}
\newcommand{\tj}{\tilde{j}}
\newcommand{\vtj}{\tilde{{\bf j}}}
\newcommand{\tJ}{{\widetilde{J}}}
\newcommand{\vk}{{\bf {k}}}
\newcommand{\tk}{\tilde{k}}
\newcommand{\vtk}{\tilde{{\bf k}}}
\newcommand{\tK}{{\widetilde{K}}}
\newcommand{\vl}{{\bf {l}}}
\newcommand{\tl}{\tilde{l}}
\newcommand{\vtl}{\tilde{{\bf l}}}
\newcommand{\tL}{{\widetilde{L}}}
\newcommand{\cJ}{{\cal J}}
\newcommand{\cK}{{\cal K}}
\newcommand{\cL}{{\cal L}}
\newcommand{\ctJ}{\widetilde{{\cal J}}}
\newcommand{\ctK}{\widetilde{{\cal K}}}
\newcommand{\ctL}{\widetilde{{\cal L}}}
\newcommand{\m}{{\vec m}}

\newcommand{\hn}{{\hat n}}
\newcommand{\vP}{{\bf \Pi}}
\newcommand{\vp}{{\bf p}}
\newcommand{\vtp}{\tilde{\bf p}}
\newcommand{\vR}{{\bf R}}

\newcommand{\hvR}{\hat {\bf R}}

\newcommand{\vZ}{{\bf Z}}

\newcommand{\Za}{{Z}^1}
\newcommand{\Zb}{\widetilde{Z}^1}
\newcommand{\Zc}{{Z}^2}
\newcommand{\Zd}{\widetilde{Z}^2}
\newcommand{\Si}{{\Sigma}}
\newcommand{\vSi}{{\bf {\Sigma}}}
\newcommand{\Int}{\displaystyle \int}

\begin{document}
\title {Restricted Gravity: Lagrangian Formalism and Explicit Solutions}
\author{S. H. Oh}
\email{shoh.physics@gmail.com}
\affiliation{Department of Consilience, Tech 
University of Korea, Siheung-Si 15073, Korea}
\author{Sangwoo Kim}
\email{sangwoo7616@gmail.com}
\affiliation{Konkuk Institute of Science and Technology,
Konkuk University, Seoul 05029, Korea}
\author{Y. M. Cho}
\email{ymcho0416@gmail.com}
\affiliation{School of Physics and Astronomy, 
Seoul National University, Seoul 08826, Korea}
\affiliation{Center for Quantum Spacetime, 
Sogang University, Seoul 04107, Korea}

\begin{abstract}
~~~~~It is well known that, making the Abelian projection of Einstein's theory one can obtain 
the restricted gravity which is simpler than Einstein's theory but describes the core dynamics 
of Einstein's gravity. In this paper we present 
the Lagrangian formalism of the restricted gravity which makes the restricted gravity a self consistent field theory by itself, independent of Einstein's theory. With this we present interesting solutions 
of the restricted gravity, in particular the gravitational cosmic string, the Bertotti-Robinson spacetime, the axisymmetric pp-wave, and 
the conformally flat waves with flat wavefront. 
Moreover, we show that in the restricted gravity
the Rosen-Bondi gravitational plane wave could be described by two spin-one Maxwellian potentials. This could play important role in 
quantum gravity. We discuss the physical implications of the restricted gravity.    
\end{abstract}
\maketitle

\section{Introduction}

The Einstein's theory of gravitation is based on 
the general invariance, which is so binding that 
it (together with the simplicity principle) strongly restricts the dynamics of gravitation. For this 
reason it has often been asserted that Einstein's theory is the simplest possible theory of gravitation allowed by the general invariance. Similar assertion has been made in gauge theory. The gauge invariance 
is so binding that it strongly restricts the theory 
to the standard non-Abelian gauge theory. 

But this assertion is an overstatement. Making 
the Abelian projection one can obtain the restricted gauge theory which has the full non-Abelian gauge symmetry yet simpler than the Yang-Mills theory. Moreover, in Einstein's theory we can also have 
the restricted gravity which has the full general invariance but simpler than Einstein's theory, which describes the core dynamics of Einstein's 
theory \cite{ijmpa09,cqg12,cqg13,gc15}.   

One way to understand this situation is to remember 
that the Einstein's theory can be viewed as a gauge 
theory of Lorentz group \cite{prd76,hehl}. Adopting 
this view and imposing the maximal Abelian isometry 
to the gravitational connection, we can decompose 
the gravitational connection and the curvature tensor 
into the restricted part of the maximal Abelian 
subgroup $H$ of Lorentz group $G$ and the valence 
part of $G/H$ component, without compromising 
the general invariance. With this we can decompose 
the Einstein's theory into the restricted part made of 
the restricted connection and the valence part made of 
the gauge covariant valence connection. This allows us 
to interpret Einstein's theory as a restricted theory 
of gravitation made of the restricted connection which 
has the valence connection as the Lorentz covariant 
gravitational source \cite{ijmpa09,cqg12,cqg13,gc15}. 

This tells that we could actually construct 
a restricted theory of gravitation which has 
the full general invariance but is much simpler 
than Einstein's theory with the restricted part 
of connection alone, which retains all topological characteristics of Einstein's theory. This implies that the restricted gravity can be interpreted to describe the core dynamics of Einstein's theory \cite{ijmpa09,cqg12,cqg13,gc15}.

The Abelian decomposition has played a crucial 
role to simplify the non-Abelian dynamics and 
reveal the important hidden structures of QCD 
that one can not find out from the conventional 
QCD \cite{prd80,plb14}. In particular, it tells 
that QCD can be viewed as the restricted QCD made 
of the restricted potential, which has the valence potential as the gauge covariant colored source. 
This has played a central role to establish 
the monopole condensation and the color confinement 
in QCD \cite{prd13,epjc19}.

Similarly the Abelian decomposition reveals 
the hidden structures of Einstein's theory and clarifies the skeleton structure of the theory 
showing that the gravitational connection and 
the curvature tensor can be decomposed to the Abelian (i.e., restricted) part and the valence part. 
As importantly it tells that the restricted part 
has the full general invariance, even though it is restricted. This implies that we could construct 
a generally invariant theory of gravitation with 
the restricted connection, which is simpler than 
the Einstein's theory. This tells that the Abelian decomposition of Einstein's theory sheds a totally 
new light on Einstein's theory of gravitation.    

So far, however, the restricted gravity has been 
shown to exist within the context of Einstein's theory. So it is important to show that the restricted gravity can exist independent of Einstein's theory,
as a self consistent theory of gravity. To do that 
we need a Lagrangian formalism of restricted 
gravity which can prove that the restricted 
gravity can indeed describe a theory of gravity. 

{\it The purpose of this paper is three-fold. First,
to construct the Lagrangian which describes the restricted gravity which can make the restricted gravity a self consistent field theory by itself. 
Second, to show that the restricted gravity indeed describes many intersting solutions, including
the gravitational cosmic string and the axisymmetric pp-waves. Third, to show that the restricted gravity can describe the Rosen-Bondi gravitational plane wave with a pair of Maxwell type spin-one gauge fields, which may have a deep implication in quantum gravity.} 

It is a non-trivial task to find the Lagrangian 
which describes the restricted gravity. The reason 
is that, in the gauge formalism of Einstein's theory 
the connection plays the fundamental role, but in Einstein's theory the metric plays the fundamental role. And in the gauge formalism of Einstein's 
theory, we have the metricity condition which translates the connection to the metric from 
the equation of motion. But in the restricted gravity we only have the restricted connection without 
the valence connection, so that it is not clear how 
the restricted connection can determine the metric. 
In this paper we show how to resolve this problem constructing the Lagrangian for the restricted gravity, which can make the restricted gravity 
a self consistent field theory.    

The remarkable feature of the restricted gravity 
is that the Abelian component of the metric is described by the Abelian gauge potential which satisfies the Maxwell type equation. This implies 
that the core dynamics of Einstein's gravity can 
be described by an Abelian gauge theory. In particular, this implies that the graviton could 
be described by a spin-one gauge field. This could have deep implication in quantum gravity, not just 
because the graviton could be identifies as 
a spin-one particle but more importantly because 
the Maxwell type gauge theory is easier to quantize. 

Since Feynman tried to quantize Einstein's theory 
quantizing the metric, the graviton is regarded as 
a spin-two particle \cite{feyn}. But this popular 
view has a critical defect, because it is well known that the graviton which couples to fermions is not 
the metric but the tetrad, four spin-one vector fields. This means that the tetrad is more fundamental than the metric, which strongly implies that 
the metric may not be viewed as the quantum field 
of gravity. The Lagrangian formalism of the restricted gravity provides an alternative way to quantize 
the gravity, with massless spin-one graviton. 

The paper is organized as follows. In Section II we review the gauge formulation of Einstein's theory for later purpose. In Section III we review the Abelian decomposition of Einstein's theory for notational purpose. In Section IV we construct the Lagrangian which describes the restricted gravity as a Maxwell type Abelian gauge theory. In Section V we discuss 
the spacetime geometry of the restricted gravity 
with trivial topology. In Section VI we present interesting solutions of the $A_2$ gravity, and 
in Section VII we present interesting solutions 
of the $B_2$ gravity. In Section VIII we discuss 
how the Rosen-Bondi gravitational wave can be 
described by a pair of spin-one gauge potentials. 
In Section IX we discuss the physical implications 
of the restricted gravity, in particular on quantum gravity.

\section{Einstein's Theory as Gauge Theory of Lorentz Group: A Review}

To discuss the restricted gravity we have to start from the gauge formalism of Einstein's theory, because the restricted gravity is obtained by 
the Abelian projection of this gauge theory of Lorentz group \cite{ijmpa09,cqg12,cqg13,gc15}. 
Let $\pd_\mu~(\mu=t,x,y,z)$ and $e_a~(a= 0,1,2,3)$ be a coordinate basis and an orthonormal basis in spacetime,
\begin{gather}
[\pro_\mu,~\pro_\nu] = 0,
~~~[e_a,~e_b] = f_{ab}^{~~c}~e_c,  \nn\\
\pro_\mu=e_\mu^{~a}~e_a,
~~~~~e_a=e_a^{~\mu} \pro_\mu,  \nn\\  
f_{ab}^{~~c}=(e_a^{~\mu} \pro_\mu e_b^{~\nu}
-e_b^{~\mu} \pro_\mu e_a^{~\nu}) e_\nu^{~c},
\label{basis}
\end{gather}
where $e_\mu^{~a}$ and $e_a^{~\mu}$ are the tetrad 
and inverse tetrad. And let $J_{ab}=-J_{ba}$ be 
the generators of Lorentz group which act on 
the tetrad,
\begin{gather}
[J_{ab}, ~J_{cd}] = \eta_{ac} J_{bd}
-\eta_{bc} J_{ad} +\eta_{bd} J_{ac}
-\eta_{ad} J_{bc}  \nn\\
=f_{ab,cd}^{~~~~~~mn}~J_{mn}, \nn\\
f_{ab,cd}^{~~~~~~mn}=\eta_{ac} \delta_b^{~[m} \delta_d^{~n]}
-\eta_{bc} \delta_a^{~[m} \delta_d^{~n]} \nn\\
+\eta_{bd} \delta_a^{~[m} \delta_c^{~n]} 
-\eta_{ad} \delta_b^{~[m} \delta_c^{~n]},
\label{lgcr}
\end{gather}
where $\eta_{ab}=diag~(-1,1,1,1)$ is the Lorentz invariant Minkowski metric. We can express $J_{ab}$
in terms of three dimensional rotation and
boost generators $L_i$ and $K_i$,
\begin{gather}
[L_i, ~L_j] = \epsilon_{ijk} L_k,
~~~[L_i, ~K_j] = \epsilon_{ijk} K_k,  \nn\\
[K_i, ~K_j] =-\epsilon_{ijk} L_k, ~~~(i,j,k= 1,2,3)  \nn\\
L_i=\dfrac12 \epsilon_{ijk} J_{jk},~~~K_i=J_{0i}, 
\end{gather} 
so that any Lorentz sextet can be decomposed 
to two three-dimensional rotation and boost 
parts. 

Now we introduce the anti-symmetric unit tensor $\vI^{ab}$ which forms an adjoint representation of Lorentz group whose rotation and boost parts $\hat m^{ab}$ and $\hat e^{ab}$ are defined by 
\begin{gather}
\vI^{ab}=\left( \begin{array}{c} {\hat m}^{ab} \\
{\hat e}^{ab} \end{array} \right), \nn\\
{\hat m}_i^{~ab}=\epsilon_{0i}^{~~ab}, 
~~~{\hat e}_i^{~ab}=\big(\delta_0^{~a} \delta_i^{~b} -\delta_0^{~b}
\delta_i^{~a} \big), \nn\\
I_{mn}^{~~~ab}=\big(\delta_m^{~a} \delta_n^{~b} -\delta_m^{~b} \delta_n^{~a} \big) 
=-(J_{mn})^{ab}.
\end{gather}
Let $\vp$ be a Lorentz sextet which forms an 
adjoint representation of Lorentz group. Clearly  
we can express $\vp$ by $p^{ab}$, 
\begin{gather}
\vp = \frac12 p_{ab} \vI^{ab}=\left( \begin{array}{c} \m \\
\e \end{array} \right),  \nn\\
p^{ab}=\vp \cdot \vI^{ab}=\frac12 p^{cd} I_{cd}^{~~~ab}, 
\label{iddef}
\end{gather}
where $m_i=\epsilon_{ijk}p^{jk}/2~(i,j,k=1,2,3)$ is the rotation (or magnetic) part and $e_i=p^{0i}$ is the boost (or electric) part 
of $\vp$. Notice that, with the invariant metric $\delta_{AB}$ of Lorentz group
\begin{gather}
\delta_{AB}= -\dfrac{1}{4} f_{AC}^{~~D}f_{BD}^{~~C}  \nn\\
=diag~(+1,+1,+1,-1,-1,-1),
\label{inme}
\end{gather}
we have
\begin{gather}
\vp^2= \delta_{AB}~p^A p^B=\dfrac 12 p_{ab} p^{ab}= \m^2-\e^2.
\end{gather}
Also, introducing the dual partner $\vtp$ of $\vp$ given by $\widetilde p^{ab}=\epsilon^{abcd} p_{cd}/2$, we have (with $\epsilon_{0123}=+1$)
\begin{gather}
\vtp=\left( \begin{array}{c} \e \\ -\m
\end{array} \right),~~~\vp \cdot \vtp
=2 \m \cdot \e,~~~\vtp^2=-\vp^2, \nn\\
\vp \times \vtp=0,~~~~~[p,~\widetilde p]=0.
\label{cl}
\end{gather}
This tells that any two vectors which are dual to 
each other are commuting.

To have the gauge formalism of Einstein's theory,  we introduce the anti-symmetric metric $g_\mn^{~~ab}$ which forms an adjoint representation of Lorentz group
\begin{gather}
\vg_\mn = g_\mn^{~~ab}~\vI_{ab}
=e_\mu^{~a} e_\nu^{~b}~\vI_{ab}, \nn\\
g_{\mu\nu}^{~~ab}=(e_\mu^{~a} 
e_\nu^{~b}-e_\nu^{~a} e_\mu^{~b})
= e_\mu^{~c} e_\nu^{~d} I_{cd}^{~~ab}.
\label{vg}
\end{gather}
To proceed, we express the gravitational connection
$\Gm_\mu^{~ab}$ corresponding to the gauge potential and the curvature tensor $R_\mn^{~~ab}$ corresponding to the gauge field strength 
the Lorentz group by Lorentz sextets $\vGm_\mu$ and $\vR_\mn$. Clearly we have
\begin{gather}
\vR_\mn=\pro_\mu \vGm_\nu-\pro_\nu \vGm_\mu
+\vGm_\mu\times \vGm_\nu. 
\end{gather}
With this we have the Lorentz invariant 
Einstein-Hilbert action of the gravitation  \cite{prd76}
\begin{gather} 
S[e_\mu^{~a},\vGm_\mu] = \dfrac1{16\pi G_N} 
\Int e~(\vg_\mn \cdot \vR^\mn)~d^4x,  \nn\\
e={\rm Det}~e_\mu^{~a}. 
\label{ehlag}
\end{gather}
This is the gauge formalism of Einstein's theory. 

Notice that the gauge formalism of Einstein's theory is not the ordinary gauge theory. The Einstein-Hilbert action is linear in $\vR_\mn$, while the Yang-Mills action is quadratic in 
the field strength. So the dynamics of Einstein's theory becomes different from the non-Abelian dynamics of the Yang-Mills theory. In particular, in Einstein's theory the metric $\vg_\mn$ (not 
the connection) becomes the dynamical variable, but in gauge theory the potential does.

In this first order formalism we have two equations of motion from (\ref{ehlag})
\begin{gather}
\delta e_\mu^{~a}:~~~~~\vg_\mn \cdot \vR^{\nu a} =0, \nn\\
\delta \vGm_\mu:~~~~~\mathscr{D}_\mu \vg^{\mu\nu}=0
~~~~(\mathscr{D}_\mu=\nabla_\mu+\vGm_\mu \times),
\label{Eeq}
\end{gather}
where $\nabla_\mu$ describes the generally covariant derivative and $\mathscr{D}_\mu$ describes the generally and Lorentz covariant derivative. The first equation assures that, in the absence of matter fields, the Ricci tensor must vanish. The second equation, with  (\ref{vg}), tells that $e_\mu^a$ must be generally and Lorentz gauge covariant constant,
\begin{gather}
\mathscr{D}_\mu e_\nu^{~a}
=\pro_\mu e_\nu^{~a}-\Gm_\mn^{~~\alpha} e_\alpha^{~a} +\Gm_{\mu b}^{~~a} e_\nu^{~b}=0,
\label{mcc}
\end{gather}
or
\begin{gather}
\Gm_\mu^{~ab}= e^{\nu a} \hat{\nabla}_{\mu} e_{\nu}^{~b}.
\label{scon}
\end{gather}
Conversely, with (\ref{mcc}) the second equation becomes an identity,
\begin{gather}
\mathscr D_\mu \vg^\mn=D_\mu \vI^{ab}=0.
\end{gather}
This tells that in the Lorentz gauge formalism of Einstein's theory the gauge potential $\vGm_\mu$ is given by the spin connection given by (\ref{scon}), which can (in principle) 
have torsion when we have the spinor matter field in the theory. This shows that the second equation is nothing but the metric compatibility of 
the connection,
\begin{gather}
\mathscr{D}_\mu \vg^\mn=0
~~~\Longleftrightarrow ~~~\nabla_\alpha g_\mn=0,
\label{mc}
\end{gather}
which assures that (\ref{ehlag}) indeed describes Einstein's theory of gravitation.

\section{Abelian Decomposition of Einstein's Theory: A Brief Review}

We can obtain the restricted gravity applying
the Abelian decomposition to the above gauge formalism of Einstein's theory \cite{ijmpa09,cqg12,cqg13,gc15}. Let $(\hn_1,\hn_2,\hn_3=\hn)$ be a $3$-dimensional unit vectors ($\hn_i^2=1$) which form a right-handed orthonormal basis of the rotation subgroup of the Lorentz group, and choose
\begin{gather}
\vl_i= \left( \begin{array}{c} \hn_i \\
0 \end{array} \right),
~~~\vk_i= \left( \begin{array}{c} 0 \\
\hn_i  \end{array} \right)= -\vtl_i. 
\label{lbasis}
\end{gather}
to be an orthonormal basis of the Lorentz group.

Since the Lorentz group has two $2$-dimensional maximal Abelian subgroups, $A_2$ whose generators are made of $L_3$ and $K_3$ and $B_2$ whose generators are made of $(L_1+K_2)/\sqrt 2$ and $(L_2-K_1)/\sqrt 2$ (or equivalently $(L_1-K_2)/\sqrt 2$ and $(L_2+K_1)/\sqrt 2$),
so that we have two possible Abelian decompositions of the gravitational connection. 
To make the Abelian decomposition, we have to select two Abelian directions $\vp$ and $\vtp$ 
in Lorentz group, and decompose the connection 
to the restricted part and the valence part, imposing the isometry which parallelizes the Abelian directions. Let $\vp$ to be one of 
the Abelian directions and let
\begin{gather}
D_\mu \vp = (\pro_\mu + \vGm_\mu \times) ~\vp=0. 
\label{ic}
\end{gather}
This automatically assures
\begin{gather}
D_\mu \vtp =(\pro_\mu + \vGm_\mu \times) ~\vtp=0, 
\label{dic}
\end{gather}
because $\epsilon_{abcd}$ is an invariant tensor. 
So, we actually need only one isometry condition (\ref{ic}). 

It is useful to characterize the isometry by two Casimir invariants. Let the isometry be described 
by $\vp$ and $\vtp$. It has two Casimir invariants $\alpha$ and $\beta$,
\begin{gather}
\alpha ={\vp \cdot \vp} =\m^2-\e^2,
~~~\beta = {\vp \cdot \vtp} =2 \m \cdot \e.
\end{gather}
But we can always choose $(\alpha,\beta)$ 
to be $(\pm 1,0)$ or $(0,0)$ unless $\alpha^2+\beta^2=0$, redefining $\vp$ and 
$\vtp$ by
\begin{gather}
\vp'=a\vp +b\vtp,
~~~\vtp'= a\vp -b\vtp.
\end{gather}
So the magnetic isometry in Einstein's theory can 
be classified by the rotation-boost (or non light-like) isometry and the null (or light-like) isometry whose Casimir invariants are denoted by 
$(\pm 1,0)$ and $(0,0)$. 

\subsection{$A_2$ (Non-Lightlike) Decomposition}

Let the maximal Abelian subgroup be $A_2$ whose isometry is made of $L_3$ and $K_3$. and let $\vp$ 
and $\vtp$ be the two isometry vector fields which correspond to $L_3$ and $K_3$. Clearly we can express the $A_2$ isometry by
\begin{gather}
\vp =\vl = \vl_3= \left(\begin{array}{c} \hn \\
0 \end{array} \right),
~~~\vtp =\vtl =\vtl_3= \left(\begin{array}{c} 0 \\
-\hn  \end{array} \right),  \nn\\
D_\mu \vl=0,~~~~D_\mu \vtl=0,
\label{a2ic}
\end{gather}
whose Casimir invariants are fixed by $(1,0)$. With this we find the restricted connection $\hvGm_\mu$ which satisfies the isometry condition
\begin{gather}
\hvGm_\mu= A_\mu ~\vl - B_\mu ~\vtl
-\vl \times \pro_\mu \vl  \nn\\
=A_\mu ~\vl - B_\mu ~\vtl
-\frac12 (\vl \times \pro_\mu \vl
-\vtl \times \pro_\mu \vtl) \nn\\
A_\mu = {\vl \cdot \vGm_\mu},
~~B_\mu  = \vtl \cdot \vGm_\mu, 
~~\vl \times \pro_\mu \vl 
= -\vtl \times \pro_\mu \vtl,
\label{a2rc}
\end{gather}
where $A_\mu$ and $B_\mu$ are two Abelian connections of $\vl$ and $\vtl$ components. 
From this we have the restricted 
field strength $\hvR_\mn$ given by
\begin{gather}
\hvR_\mn=\pro_\mu \hvGm_\nu-\pro_\nu \hvGm_\mu
+\hvGm_\mu \times \hvGm_\nu  \nn\\
=(A_\mn+ H_\mn)~\vl -(B_\mn+\tilde H_\mn)~\vtl  \nn\\
=\bar A_\mn~\vl -B_\mn~\vtl, \nn\\
\bar A_\mn=A_\mn+ H_\mn=\pro_\mu \bar A_\nu 
- \pro_\nu \bar A_\mu,  \nn\\
A_\mn = \pro_\mu A_\nu - \pro_\nu A_\mu,  \nn\\
H_\mn = -\vl \cdot (\pro_\mu \vl \times \pro_\nu \vl)
= \pro_\mu C_\nu - \pro_\nu C_\mu,  \nn\\
\bar A_\mu= A_\mu +C_\mu,
~~~~~C_\mu= -\hn_1 \cdot \pro_\mu \hn_2,  \nn\\
B_\mn= \pro_\mu B_\nu - \pro_\nu B_\mu,  \nn\\
\tilde H_\mn = -\vtl \cdot (\pd_\mu \vl \times \pd_\nu \vl) 
= \vtl \cdot (\pd_\mu \vtl \times \pd_\nu \vtl) 
= 0.
\label{a2rct}
\end{gather}
Notice that here $\tilde H_\mn$ vanishes, so that only $\bar A_\mu$ and $\bar A_\mn$ has a dual structure. In other words, the boost part (i.e., the $\vtl$ component) of $\hvR_\mn$ has no topological part. Moreover, $C_\mu$ and $H_\mn$ 
in $\bar A_\mu$ and $\bar A_\mn$ are formally 
identical to the magnetic potential and field strength of $SU(2)$ QCD. This tells that 
the topology of the Einstein's theory is fixed 
by the topology of the $SU(2)$ subgroup, 
which is identical to that of SU(2) QCD.

With this we obtain the full connection of Lorentz group adding the Lorentz covariant valence connection $\vZ_\mu$, 
\begin{gather}
\vGm_\mu = \hvGm_\mu + \vZ_\mu, \nn\\
\vZ_\mu = \Za_\mu \vl_1 -\Zb_\mu \vtl_1
+\Zc_\mu \vl_2 -\Zd_\mu \vtl_2.
\label{a2z}
\end{gather} 
The corresponding field strength $\vR_\mn$ which describes the curvature tensor is decomposed to 
\begin{gather}
\vR_\mn = \hvR_\mn+\vZ_\mn, \nn\\
\vZ_\mn=\hD_\mu \vZ_\nu - \hD_\nu \vZ_\mu
+ \vZ_\mu \times \vZ_\nu,
\end{gather}
where $\hD_\mu=\pro_\mu + \hvGm_\mu \times$ and $\vZ_\mn$ is the valence part of the curvature tensor. This is the $A_2$ decomposition of the gravitational connection. 

Now, we decompose the metric $\vg_\mn$ to the restricted and valence parts,
\begin{gather}
\vg_\mn = \hvg_\mn + \vG_\mn,  \nn\\
\hvg_\mn =e_\mu^a~e_\nu^b~\vSi_{ab}=G_\mn ~\vl -\tG_\mn \vtl, \nn\\
\vSi_{ab}=l_{ab}~\vl-\tl_{ab}~\vtl, \nn\\
G_\mn=e_\mu^a~e_\nu^b~l_{ab},
~~~\tG_\mn = e_\mu^a~e_\nu^b~\tl_{ab}, \nn\\
\vG_\mn= e_\mu^a~e_\nu^b~\vP_{ab}=G_\mn^1 \vl_1-\tG_\mn^1 \vtl_1 
+G_\mn^2 \vl_2-\tG_\mn^2 \vtl_2, \nn\\
\vP_{ab}=\vI_{ab}-\vSi_{ab}
=l^1_{ab}~\vl_1-\tl^1_{ab}~\vtl_1
+l^2_{ab}~\vl_2-\tl^2_{ab}~\vtl_2, \nn\\
G_\mn^1=e_\mu^a~e_\nu^b~l_{ab}^1,
~~~\tG_\mn^1 = e_\mu^a~e_\nu^b~\tl_{ab}^1, \nn\\
G_\mn^2=e_\mu^a~e_\nu^b~l_{ab}^2,
~~~\tG_\mn^2 = e_\mu^a~e_\nu^b~\tl_{ab}^2,  \nn\\
\vZ_\mu \cdot \vSi_{ab}=0,
~~~\vZ_\mu \cdot \vP^{ab}=Z_\mu^{~ab}.
\label{gdeca}
\end{gather}
Notice that   
\begin{gather}
\Si_{ab}^{~~cd}~\vSi_{cd}= \vSi_{ab},  
~~~~\Pi_{ab}^{~~cd}~\vP_{cd} = \vP_{ab}, \nn\\
\vSi_{ab} \cdot \vP^{cd} =0.
\end{gather}
So $\vSi_{ab}$ and $\vP_{ab}$ can be viewed as 
the projection operators which project the metric to the restricted and valence parts.

With this we can express the Einstein-Hilbert Lagrangian as
\begin{gather}
{\cal L}=\dfrac{e}{16\pi G_N} \Big[~(\hvg_\mn 
+ \vG_\mn) \cdot (\hvR_\mn+\vZ_\mn) \Big] \nn\\
=\dfrac{e}{16\pi G_N} ~\Big[~\hvg_\mn
\cdot \hvR_\mn +\hvg_\mn \cdot \vZ_\mn  
+\vG_\mn \cdot \vZ_\mn \Big],
\label{a2dlag}
\end{gather}
from which we get the following equations of motion
\begin{gather}
\del e_\mu^a:~G_\mn (\cD^\nu \bA^\rho-\cD^\rho \bA^\nu)
-\tG_\mn (\cD^\nu B^\rho-\cD^\rho B^\nu) \nn\\
+\vG_\mn \cdot (\hD^\nu \vZ^\rho-\hD^\rho \vZ^\nu)=0, \nn\\
\del A_\mu:~\nabla_\mu G^\mn 
+\vl \cdot (\vZ_\mu \times \vG^\mn) =0, \nn\\
\del B_\mu:~\nabla_\mu \tG^\mn 
+\vtl \cdot (\vZ_\mu \times \vG^\mn) =0, \nn\\
\del \vZ_\mu:~\hat{{\mathscr D}}_\mu \vG^\mn 
+\vZ_\mu \times \hvg^\mn=0.
\label{a2Eeq2}
\end{gather}
This is the equation for the $A_2$ decomposition of Einstein's theory. The first equation is identical to the first equation of (\ref{Eeq}), and the next three equations can be combined to 
a single equation $\mathscr D_\mu \vg^\mn=0$. This assures that the above equation is identical to the Einstein's equation (\ref{Eeq}). Another point is that the last equation $\hat{{\mathscr D}}_\mu \vG^\mn +\vZ_\mu \times \hvg^\mn=0$ can be replaced by 
\begin{gather}
\hat{{\mathscr D}}_\mu \hvg^\mn 
+\vZ_\mu \times \vG^\mn=0,
\end{gather}
because $\mathscr D_\mu \vg^\mn=0$ makes them equivalent. 

\subsection{$B_2$ (Light-like) Decomposition}

Now, let the Abelian subgroup is $B_2$ whose 
isometry generators is given by $(L_1+K_2)/\sqrt 2$ and $(L_2-K_1)/\sqrt 2$, and introduce 
a complete basis made of four null vectors $\vj$, $\vtj$, $\vk$, $\vtk$, and $\vl$, $\vtl$
\begin{gather}
\vj =\frac{\vl_1+ \vk_2}{\sqrt 2}, 
~~~\vtj =\frac{\vl_2-\vk_1}{\sqrt 2}, \nn\\
\vk= \frac{\vl_1 -\vk_2}{\sqrt 2},
~~~\vtk= -\frac{\vl_2 +\vk_1}{\sqrt 2}.  
\end{gather}
With this we can express the $B_2$ isometry  
by $\vj$ and $\vtj$ whose Casimir invariant 
is given by $(0,0)$.

In this case we can express the $B_2$ isometry by
\begin{gather}
\vj =\frac{\vl_1+ \vk_2}{\sqrt 2}
=\frac{1}{\sqrt 2} \left( \begin{array}{c} \hn_1 \\
\hn_2  \end{array} \right), \nn\\
\vtj =\frac{\vl_2-\vk_1}{\sqrt 2}
=\frac{1}{\sqrt 2} \left( \begin{array}{c}  \hn_2 \\ -\hn_1  \end{array} \right), \nn\\
D_\mu \vj=0,~~~D_\mu \vtj=0.
\label{b2ic}
\end{gather}
With this we find the following restricted connection for the $B_2$ gravity,
\begin{gather}
\hvGm_\mu =\Gm_\mu~\vj -\tGm_\mu~\vtj 
-\vk \times \pd_\mu \vj  \nn\\
= \Gm_\mu~\vj - \tGm_\mu~\vtj
- \dfrac12(\vk \times \pd_\mu \vj
-\vtk \times \pd_\mu \vtj),  \nn\\
\Gm_\mu = \vk \cdot \vGm_\mu,
~~~\tGm_\mu = \vtk \cdot \vGm_\mu, 
\label{b2rc}
\end{gather}
where $\Gamma_\mu$ and $\tGm_\mu$ are two Abelian connections of $\vj$ and $\vtj$ components.

The restricted curvature tensor $\hvR_\mn$ is given by
\begin{gather}
\hvR_\mn=\pd_\mu \hvGm_\nu-\pd_\nu \hvGm_\mu
+\hvGm_\mu \times \hvGm_\nu 
=K_\mn \vj-\tK_\mn \vtj,  \nn\\
K_\mn=\Gm_\mn+H_\mn=\pd_\mu K_\nu 
-\pd_\nu K_\mu,  \nn\\
\tK_\mn=\tGm_\mn+\tH_\mn=\pd_\mu \tK_\nu 
-\pd_\nu \tK_\mu, \nn\\
\Gm_\mn =\pd_\mu \Gm_\nu -\pd_\nu \Gm_\mu,
~~~\tGm_\mn =\pd_\mu \tGm_\nu -\pd_\nu \tGm_\mu, \nn\\
H_\mn = -\vk \cdot (\pd_\mu \vj \times \pd_\nu \vk
-\pd_\nu \vj \times \pd_\mu \vk)  
= \pd_\mu C_\nu^1 - \pd_\nu C_\mu^1,  \nn\\
C_\mu^1=-\frac{1}{\sqrt 2} \hn_2\cdot \pd_\mu \hn_3,  \nn\\
\tH_\mn= -\vtk \cdot (\pd_\mu \vj \times \pd_\nu \vk -\pd_\nu \vj \times \pd_\mu \vk)
=\pd_\mu C_\nu^2 - \pd_\nu C_\mu^2,  \nn\\
C_\mu^2=\frac{1}{\sqrt 2} \hn_3\cdot \pd_\mu \hn_1,  \nn\\
K_\mu=\Gm_\mu+C_\mu^1,
~~~~~\tK_\mu=\tGm_\mu+C_\mu^2.
\label{b2rct}
\end{gather}
Notice that here, unlike the $A_2$ gravity, both 
$K_\mn$ and $\tK_\mn$ have the topological parts $H_\mn$ and $\tH_\mn$. Another point is that, $\hvR_\mn$ is orthogonal to $\vl$ and $\vtl$. 
This should be contrasted with the restricted curvature tensor (\ref{a2rct}) of the $A_2$ gravity. With this we obtain the $B_2$ decomposition of the connection and curvature tensor,
\begin{gather}
\vGm_\mu = \hvGm_\mu + \vZ_\mu,  \nn\\
\vZ_\mu = J_\mu \vk -\tJ_\mu \vtk +L_\mu \vl -\tL_\mu \vtl,  \nn\\
\vR_\mn =\hvR_\mn+\vZ_\mn, \nn\\
\vZ_\mn=\hD_\mu \vZ_\nu - \hD_\nu \vZ_\mu
+ \vZ_\mu \times \vZ_\nu, 
\end{gather}
\label{b2z}
where $\vZ_\mu$ and $\vZ_\mn$ are the valence parts and $\hD_\mu=\pro_\mu + \hvGm_\mu \times$. 

For the $B_2$ decomposition we may adopt the following decomposition of the metric $\vg_\mn$ 
\begin{gather}
\vg_\mn = \hvg_\mn +\vG_\mn, \nn\\
\hvg_\mn =e_\mu^a~e_\nu^b~\vSi_{ab}   
=\cJ_\mn~\vk-\ctJ_\mn~\vtk
+\cK_\mn~\vj-\ctK_\mn~\vtj, \nn\\
\vSi_{ab}=j_{ab}~\vj-\tj_{ab}~\vtj
+k_{ab}~\vk-\tk_{ab}~\vtk, \nn\\
\cJ_\mn=e_\mu^a~e_\nu^b~j_{ab},
~~~\ctJ_\mn=e_\mu^a~e_\nu^b~\tj_{ab}, \nn\\
\cK_\mn=e_\mu^a~e_\nu^b~k_{ab},
~~~\ctK_\mn=e_\mu^a~e_\nu^b~\tk_{ab}, \nn\\
\vG_\mn= e_\mu^a~e_\nu^b~\vP_{ab} 
= \cL_\mn~\vl-\ctL_\mn~\vtl,  \nn\\
\cL_\mn=e_\mu^a~e_\nu^b~l_{ab},
~~~\ctL_\mn=e_\mu^a~e_\nu^b~\tl_{ab},  \nn\\
\vP_{ab}=\vI_{ab}-\vSi_{ab} 
= l_{ab}~\vl-\tl_{ab}~\vtl.
\label{gdecb}
\end{gather}
Notice that this decomposition of the metric is different from (\ref{gdeca}). In fact here $\vG_\mn$ plays the role of $\hvg_\mn$ (and $\vP_{ab}$ becomes $\vSi_{ab}$) in (\ref{gdeca}). But here again we have 
\begin{gather}
\Si_{ab}^{~~cd}~\vSi_{cd}= \vSi_{ab},  
~~~~\Pi_{ab}^{~~cd}~\vP_{cd} = \vP_{ab}, \nn\\
\vSi_{ab} \cdot \vP^{cd} =0.
\end{gather}
so that $\vSi_{ab}$ and $\vP_{ab}$ become projection operators.

With this we have the $B_2$ decomposition of Einstein-Hilbert Lagrangian, 
\begin{gather}
\cL =\dfrac{e}{16\pi G_N} \Big[~(\hvg_\mn 
+ \vG_\mn) \cdot (\hvR_\mn+\vZ_\mn) \Big] \nn\\
=\dfrac{e}{16\pi G_N} ~\Big[~\hvg_\mn
\cdot \hvR_\mn +\hvg_\mn \cdot \vZ_\mn  
+\vG_\mn \cdot \vZ_\mn \Big].
\label{b2dlag}
\end{gather}
From this we have
\begin{gather}
\del e_\mu^a:~\cJ_\mn (\cD^\nu K^\rho-\cD^\rho K^\nu)
-\ctJ_\mn (\cD^\nu \tK^\rho-\cD^\rho \tK^\nu) \nn\\
+\cK_\mn (\cD^\nu J^\rho-\cD^\rho J^\nu)
-\ctK_\mn (\cD^\nu \tJ^\rho-\cD^\rho \tJ^\nu) \nn\\
+\vG_\mn \cdot \vZ^{\nu\rho} =0, \nn\\
\del \Gm_\mu:~\nabla_\mu \cJ^\mn 
+\vj \cdot (\vZ_\mu \times \vg^\mn)=0, \nn\\
\del \tGm_\mu:~\nabla_\mu \ctJ^\mn 
+\vtj \cdot (\vZ_\mu \times \vg^\mn)=0, \nn\\
\del \vZ_\mu:~\hat{{\mathscr D}}_\mu \vG^\mn 
+(\nabla_\mu \cK^\mn~\vj- \nabla_\mu\ctK^\mn~\vtj) \nn\\
+\vZ_\mu \times (\cK^\mn~\vj-\ctK^\mn~\vtj) 
+ (K_\mu \vj - \tK_\mu \vtj) \times \hvg^{\mn} = 0.
\label{b2Eeq2}
\end{gather}
This is the equation for the $B_2$ decomposition of Einstein's theory. And here again we can combine the last three equations to a single equation $\mathscr D_\mu \vg^\mn=0$. This assures 
that the above (\ref{b2Eeq2}) is identical to 
the Einstein's equation (\ref{Eeq}). Moreover, 
the last equation of (\ref{b2Eeq2}) (with $\mathscr D_\mu \vg^\mn=0$) can be replaced by 
\begin{gather}
\hat{{\mathscr D}}_\mu \hvg^{\mn} 
-(\nabla_\mu \cK^\mn~\vj
-\nabla_\mu \ctK^\mn~\vtj ) -(K_\mu \vj - \tK_{\mu} \vtj ) \times \hvg^\mn  \nn\\
+ \vZ_\mu \times (\cJ^\mn \vk -\ctJ^\mn \vtk 
+ \vG^\mn) = 0. 
\end{gather}
Of course, we could make the mathematically identical $B_2$ decomposition with $\vk$ and $\vtk$ as the isometry.

\section{Restricted Gravity: Lagrangian Formalism}

The decomposed equations (\ref{a2Eeq2}) and (\ref{b2Eeq2}) are not just identical to 
the Einstein's equation. They contain more 
information, in the sense that they reveal 
the hidden structure of Einstein's equation. 
This must be clear because they show that 
the Einstein's equation is made of two parts, 
the restricted part which describes the Abelian subdynamics and the valence part which plays 
the role of the gravitational source. This
tells us two things. First, Einstein's theory can 
be simplified to the restricted gravity which we 
can obtain putting $\vZ_\mu=0$. Second, Einstein's 
theory can be reinterpreted as the restricted 
gravity which has the gauge covariant valence connection as the extra source.
 
This is nice, but this does not guarantee that 
the restricted gravity is a self-consistent field theory by itself. To show that the restricted 
gravity exists independent of Einstein's theory 
we need to have the Lagrangian which can describe 
the restricted gravity. To construct such a Lagrangian, however, is a non-trivial task. This is because in 
the gauge formalism of Einstein's theory we have 
the equation of motion which determines the metric 
from the connection. On the other hand, the restricted gravity is described by the restricted connection, 
so that it is unclear how we can obtain the equation 
of motion which can determine the metric from 
the restricted connection. In the following we discuss how to resolve this problem and obtain the Lagrangian formalism of the restricted gravity.

\subsection{Lagrangian Formalism of $A_2$ Gravity}

From (\ref{a2dlag}) we might naively choose 
the Lagrangian for the $A_2$ gravity to be \cite{gc15}
\begin{gather}	
\cL_A =\dfrac{e}{16\pi G_N}~\big(~\hvg_\mn \cdot \hvR_\mn \big).
\label{a2lagx}
\end{gather}
But as we have pointed out this does not work, 
because here the restricted connection can not determine the restricted metric. More seriously, 
we need to have the full metric for the spacetime. This clearly tells that the above Lagrangian 
is unacceptable. This shows that it is 
a non-trivial matter to have a Lagrangian formalism of the restricted gravity. 

Fortunately we can obtain the Lagrangian for 
the $A_2$ gravity from (\ref{a2lag1}) with 
the constraint for the $A_2$ Abelian projection, 
\begin{gather}
\cL_A =\dfrac{e}{16\pi G_N}~\Big[\vg_\mn \cdot \vR_\mn  
+ \vlam_\mu \cdot D_\mu \vl \Big] \nn\\
=\dfrac{e}{16\pi G_N} ~\Big[~\hvg_\mn \cdot \hvR_\mn 
+\hvg_\mn \cdot \vZ_\mn +\vG_\mn \cdot \vZ_\mn \nn\\
+ \vlam_\mu \cdot D_\mu \vl \Big].
\label{a2lag}
\end{gather}
Here the constraint $D_\mu \vl=0$ (with $\vl^2=1$) tells that the Lagrangian is for the $A_2$ gravity. 

To show that the Lagrangian describes the $A_2$ gravity, notice that we have the following equation of motion from the first expression 
of the above Lagrangian
\begin{gather}
\del e_\mu^a:~G_\mn (\cD^\nu \bA^\rho-\cD^\rho \bA^\nu)
-\tG_\mn (\cD^\nu B^\rho-\cD^\rho B^\nu) \nn\\
+\vG_\mn \cdot (\hD^\nu \vZ^\rho-\hD^\rho \vZ^\nu)
-\frac12 \big(\vlam^{\nu} \cdot 
D_{\nu} \vl \big)~\del_\mu^{~\rho} =0, \nn\\
\del \vGm_\mu:~D_{\mu} \vg^{\mn} 
+ \frac12 \big(\vlam^{\nu} \times \vl \big) =0, \nn\\
\del \vlam_\mu:~D_\mu \vl=0.
\label{a2Req1}
\end{gather}
One might wonder why we do not make $\del \vl$ to obtain the equation for $\vl$. This is not needed because it is a gauge degree. Indeed with $\del \vl$ we do not obtain any new information.

To simplify the above equation we notice that 
the third equation is nothing but the isometry condition which tells
\begin{gather}
\vGm_\mu = \hvGm_\mu \qquad \textrm{or} \qquad
\vZ_\mu = 0.
\label{isom}
\end{gather}
Using this in the second equation, we have
\begin{gather}
\vlam^\nu = \lam^\nu \vl - \tilde \lam^\nu \vtl 
- 2 \vl \times \hat D_\mu \vG^\mn,
\label{lamb}
\end{gather}
where $\lam^\nu = \vlam^\nu \cdot \vl$ and 
$\tilde \lam^\nu = \vlam^\nu \cdot \vtl$. With this the second equation can be expressed by 
\begin{gather}
\hat D _\mu \hvg^\mn = 0,
\end{gather}
or componentwise
\begin{gather}
\nabla_\mu G^\mn =0,~~~~~\nabla_\mu \tG^\mn =0.
\end{gather}
Also, with (\ref{isom}) the first equation is expressed by
\begin{gather}
G_\mn (\pro^\nu \bA^\rho-\pro^\rho \bA^\nu)
-\tG_\mn (\pro^\nu B^\rho-\pro^\rho B^\nu) =0.
\end{gather}
So we can express (\ref{a2Req1}) by
\begin{gather}
G_\mn (\pro^\nu \bA^\rho-\pro^\rho \bA^\nu)
-\tG_\mn (\pro^\nu B^\rho-\pro^\rho B^\nu) =0, \nn\\
\nabla_\mu G^\mn =0,~~~~~\nabla_\mu \tG^\mn =0,  \nn\\
\vZ_\mu=0.
\label{a2Req0}
\end{gather}
This is the equation for $A_2$ gravity. Notice that 
this is exactly the same equation that we obtain from 
(\ref{a2Eeq2}) with $\vZ_\mu=0$, except that here we 
have $\vZ_\mu=0$ as an independent equation.

We could derive the same equation from the second expression of (\ref{a2lag}),
\begin{gather}
\del e_\mu^a:~G_\mn (\cD^\nu \bA^\rho-\cD^\rho \bA^\nu)
-\tG_\mn (\cD^\nu B^\rho-\cD^\rho B^\nu) \nn\\
+\vG_\mn \cdot (\hD^\nu \vZ^\rho-\hD^\rho \vZ^\nu)=0, \nn\\
\del A_\mu:~\nabla_\mu G^\mn 
+\vl \cdot (\vZ_\mu \times \vG^\mn) =0, \nn\\
\del B_\mu:~\nabla_\mu \tG^\mn 
+\vtl \cdot (\vZ_\mu \times \vG^\mn) =0, \nn\\
\del \vZ_\mu:~\hat{{\mathscr D}}_\mu \hvg^\mn 
+\vZ_\mu \times \vG^\mn=0,  \nn\\
\del \vlam_\mu:~D_\mu \vl=0.
\label{a2fulleqvac}
\end{gather}
Since the last equation means $\vZ_\mu=0$, 
we can reduce the above equation to
\begin{gather}
G_\mn (\pro^\nu \bA^\rho-\pro^\rho \bA^\nu)
-\tG_\mn (\pro^\nu B^\rho-\pro^\rho B^\nu) =0, \nn\\
\nabla_\mu G^\mn =0,~~~~~\nabla_\mu \tG^\mn =0,  \nn\\
\hat{\mathscr D}_\mu \hvg^\mn=0,   \nn\\
\vZ_\mu=0.
\label{a2Req2}
\end{gather}
But the equation $\hat{\mathscr D}_\mu \hvg^\mn=0$ becomes 
redundant, because this is equivalent to the two equations 
in the second line. So (\ref{a2Req2}) reduces to (\ref{a2Req0}). 
This assures that the two expressions in (\ref{a2lag})
are indeed equivalent.

Notice that the restricted $A_2$ Lagrangian (\ref{a2lag}) can also be expressed by
\begin{gather}
\cL_A =\dfrac{e}{16\pi G_N}~\Big[~\hvg_\mn \cdot \hvR_\mn + \vlam_\mu \cdot \vZ_\mu \Big].
\label{a2lag1}
\end{gather}
Indeed, with this we have
\begin{gather}
\del e_\mu^a:~(e_\mu^a e_\nu^b)[(\pd^\nu \bar A^\rho 
- \pd^\rho \bar A^\nu )l_{ab}   \nn\\
- (\pd^\nu B^\rho - \pd^\rho B^\nu )\tl_{ab}] e_{\rho c} = 0 \nn\\
\del A_\mu:~\nabla_\mu (e^\mu_a e^\nu_b l_{ab}) =0,   \nn\\
\del B_\mu:~\nabla_\mu (e^\mu_a e^\nu_b \tl_{ab}) =0, \nn\\
\del \vlam_\mu:~\vZ_\mu=0, 
\end{gather}
which is identical to (\ref{a2Req0}).

At this point one might wonder why we can not describe 
the restricted gravity with only the restricted metric $\hvg_\mn$ and the restricted curvature $\hvR_\mn$. Indeed, with the first term in (\ref{a2lag1}) we have the first two lines of equation (\ref{a2Req2}), 
which is precisely the full Einstein's equation (\ref{a2Req1}) which satisfies the constraint 
$\vZ_\mu=0$. If so, why do we need the second term 
in (\ref{a2lag1}) which produces the last equation $\vZ_\mu=0$ which is completely independent of 
the other equations. 

To understand the reason we should keep in mind that we have constructed the restricted gravity viewing 
the Einstein's theory as a gauge theory of Lorentz group and making the valence connection vanishing. 
In this gauge formalism the connection $\vGm_\mu$ becomes a fundamental field. On the other hand, Einstein's theory is the theory of spacetime where 
the metric $\vg_\mn$, not the connection $\vGm_\mu$, plays the fundamental role. And the bridge which translates the two views to each other is the metric compatibility condition $\mathscr{D}_\mu \vg_\mn=0$. 
In this translation a vanishing connection does not mean a vanishing metric, but a flat metric. 
In fact we can not have a vanishing metric in spacetime, because the metricity (that the spacetime 
is locally Minkowskian) is the fundamental assumption of Einstein's theory. This means that, unlike in ordinary gauge theory, here we have to keep the condition $\vZ_\mu=0$ in the equation of motion, to recover the metric with the metric compatibility condition. This is why we can not describe 
the restricted gravity with the restricted metric 
and restricted connection alone. We need the full metric compatibility condition.

The physical meaning behind the restricted gravity 
is really surprising. Notice that the first equation 
of (\ref{a2Req0}) is a first-order differential equation, so that it does not describe the dynamical (i.e., propagating) graviton. It is the constraint equation which determines the connection in terms 
of the restricted metric. But remarkably the two 
equations for $G_\mn$ and $\tG_\mn$ in the second 
line look like the free Maxwell's equations. 
Indeed, since $G_\mn$ and $\tG_\mn$ are dual to 
each other, we can express $G_\mn$ by one-form potential $G_\mu$
\begin{gather}
G_\mn=\nabla_\mu G_\nu -\nabla_\nu G_\mu
= \pro_\mu G_\nu -\pro_\nu G_\mu,  
\label{a2pot1}
\end{gather}
using the fact $\nabla_\mu \tG^\mn=0$. Equivalently, we can express $\tG_\mn$ by one-form potential 
$\tilde G_\mu$
\begin{gather}
\tG_\mn=\nabla_\mu \tilde G_\nu 
-\nabla_\nu \tilde G_\mu
= \pro_\mu \tilde G_\nu -\pro_\nu \tilde G_\mu,  
\label{a2pot2}
\end{gather}
using the fact $\nabla_\mu G^\mn=0$. 

This tells that we can express the equations of 
the restricted metric $G_\mn$ and $\tG_\mn$ 
(i.e., the Abelian component of the metric) as 
a Maxwell type second-order differential equation 
in terms of the potential $G_\mu$,
\begin{gather}
\nabla_\mu G^\mn=0,~~~~~G_\mn= \pro_\mu G_\nu 
-\pro_\nu G_\mu.
\label{a2Req4}
\end{gather}
This is really remarkable, because this shows that 
the restricted metric of $A_2$ gravity can be described by an Abelian spin-one gauge potential. 
On the other hand, we emphasize that here 
the equation of motion which determine this 
restricted metric, i.e., the first equation of (\ref{a2Req0}), is different from that of 
the ordinary gauge theory.

\subsection{Lagrangian Formalism of $B_2$ Gravity}
 
The above analysis suggests that the Lagrangian 
for the $B_2$ gravity can be expressed by 
\begin{gather}
\cL =\dfrac{e}{16\pi G_N} \Big[\vg_\mn \cdot \vR_\mn 
+\vlam_\mu \cdot D_\mu \vj \Big]   \nn\\
=\dfrac{e}{16\pi G_N} ~\Big[\hvg_\mn \cdot \hvR_\mn 
+\hvg_\mn \cdot \vZ_\mn 
+\vG_\mn \cdot \vZ_\mn \nn\\
+ \vlam_\mu \cdot D_\mu \vj \Big].
\label{b2lag}
\end{gather}
Here the constraint $D_\mu \vj=0$ (with $\vj=0$) 
tells that the Lagrangian is for the $B_2$ gravity. 

From the first expression we have
\begin{gather}
\del e_\mu^a:~\cJ_\mn (\cD^\nu K^\rho-\cD^\rho K^\nu)
-\ctJ_\mn (\cD^\nu \tK^\rho-\cD^\rho \tK^\nu) \nn\\
+\cK_\mn (\cD^\nu J^\rho-\cD^\rho J^\nu)
-\ctK_\mn (\cD^\nu \tJ^\rho-\cD^\rho \tJ^\nu) \nn\\
+\vG_\mn \cdot \vZ^{\nu\rho} - \frac12 
\big(\vlam^\nu \cdot D_\nu \vj \big)~\del_\mu^{~\rho}=0, \nn\\
\del \vGm_\mu:~D_\mu \vg^\mn 
+\frac12 \big(\vlam^\nu \times \vj \big) = 0, \nn\\
\del \vlam_\mu:~D_\mu \vj=0.
\label{b2Req1}
\end{gather}
Here again we do not need $\del \vj$ to have the equation 
for $\vj$. To simplify the above equation we notice that the third equation is nothing but the isometry 
condition 
\begin{gather}
\vGm_\mu = \hvGm_\mu \qquad \textrm{or} \qquad
\vZ_\mu = 0.
\label{isob2}
\end{gather}
Using this in the second equation, we have
\begin{gather}
\vlam^\nu = \lam^\nu \vl -\tilde \lam^\nu \vtl 
- 2 \vk \times \hat D_\mu \vg^\mn,
\label{lamb2}
\end{gather}
where $\lam^\nu = \vlam^\nu \cdot \vk$ and 
$\tilde \lam^\nu = \vlam^\nu \cdot \vtk$. With this 
the second equation can be expressed by
\begin{gather}
\hat D_\mu \hvg^\mn 
+\vj \times (\vk \times \hat D_\mu \hvg^\mn)= 0,
\end{gather}
or equivalently
\begin{gather}
\nabla_\mu \cJ^\mn =0,~~~~~\nabla_\mu \ctJ^\mn =0.
\end{gather}
Also, using the isometry condition (\ref{isob2}) 
in the first equation, we have
\begin{gather}
G_\mn (\pro^\nu \bA^\rho-\pro^\rho \bA^\nu)
-\tG_\mn (\pro^\nu B^\rho-\pro^\rho B^\nu) =0.
\end{gather}
So (\ref{b2Req1}) can be expressed by 
\begin{gather}
\cJ_\mn (\pro^\nu K^\rho-\pro^\rho K^\nu)
-\ctJ_\mn (\pro^\nu \tK^\rho-\pro^\rho \tK^\nu) =0, \nn\\
\nabla_\mu \cJ^\mn =0,~~~~~\nabla_\mu \ctJ^\mn =0,  \nn\\
\vZ_\mu=0.
\label{b2Req0}
\end{gather}
This is the equation for the $B_2$ gravity. 
Here again this is exactly the same equation that we obtain from (\ref{b2Eeq2}) with $\vZ_\mu=0$, except that here we have $\vZ_\mu=0$ as an independent equation.

We can obtain the same equation of motion from the second expression of (\ref{b2lag}). From the second expression we have 
\begin{gather}
\del e_\mu^a:~\cJ_\mn (\cD^\nu K^\rho-\cD^\rho K^\nu)
-\ctJ_\mn (\cD^\nu \tK^\rho-\cD^\rho \tK^\nu) \nn\\
+\cK_\mn (\cD^\nu J^\rho-\cD^\rho J^\nu)
-\ctK_\mn (\cD^\nu \tJ^\rho-\cD^\rho \tJ^\nu) \nn\\
+\vG_\mn \cdot \vZ^{\nu\rho} =0,  \nn\\
\del \Gm_\mu:~\nabla_\mu \cJ^\mn 
+\vj \cdot (\vZ_\mu \times \vg^\mn)=0, \nn\\
\del \tGm_\mu:~\nabla_\mu \ctJ^\mn 
+\vtj \cdot (\vZ_\mu \times \vg^\mn)=0, \nn\\
\del \vZ_\mu:~\hat{{\mathscr D}}_\mu \vG^\mn 
+(\nabla_\mu \cK^\mn~\vj
- \nabla_\mu\ctK^\mn~\vtj) \nn\\
+\vZ_\mu \times (\cK^\mn~\vj-\ctK^\mn~\vtj) 
+ (K_\mu \vj - \tK_\mu \vtj) \times \hvg^{\mn} = 0.
\label{b2fulleqvac}
\end{gather}
Using the last equation we can express this by
\begin{gather}
\cJ_\mn (\pro^\nu K^\rho-\pro^\rho K^\nu)
-\ctJ_\mn (\pro^\nu \tK^\rho
-\pro^\rho \tK^\nu) =0, \nn\\
\nabla_\mu \cJ^\mn =0,
~~~~~\nabla_\mu \ctJ^\mn =0,  \nn\\
\hat{{\mathscr D}}_\mu \vG^\mn 
+(\nabla_\mu \cK^\mn~\vj
-\nabla_\mu \ctK^\mn~\vtj) \nn\\
+(K_\mu \vj -\tK_\mu \vtj) \times \hvg^{\mn} =0, \nn\\
\vZ_\mu=0.
\label{b2Req2}
\end{gather}
But here again the last equation is redundant 
because it is equivalent to the two equations in 
the second line. So the above equation becomes identical to (\ref{b2Req0}). This assures that 
the two expressions in (\ref{b2lag}) are 
indeed equivalent.

Notice that, just as in the $A_2$ gravity, the $B_2$ 
gravity can also be described by
\begin{gather}
\cL =\dfrac{e}{16\pi G_N} \Big[\hvg_\mn \cdot \hvR_\mn 
+\vlam_\mu \cdot \vZ_\mu \Big].  
\label{b2rg2}
\end{gather}
Here again the second term plays the crucial role to
assure the full metric compatibility of the connection 
to recover the metric from the connection.    

The physics of (\ref{b2Req0}) is clear. Here again 
the first equation can be viewed as the constraint 
equations which determine the connection in terms 
of the restricted metric. But the two equations for $\cJ_\mn$ and $\ctJ_\mn$ allow us to introduce one-form potential $\cJ_\mu$ for $\cJ_\mn$
\begin{gather}
\cJ_\mn= \pro_\mu \cJ_\nu -\pro_\nu \cJ_\mu,  
\label{b2pot1}
\end{gather}
or $\ctJ_\mn=  \pro_\mu \ctJ_\nu -\pro_\nu \ctJ_\mu$. With this we can express the equations of 
the restricted metric $\cJ_\mn$ and $\ctJ_\mn$, 
the Abelian component of the metric, as 
a Maxwell type second order differential equation 
in terms of the potential $\cJ_\mu$,
\begin{gather}
\nabla_\mu \cJ^\mn=0,~~~~~\cJ_\mn
= \pro_\mu \cJ_\nu -\pro_\nu \cJ_\mu.
\label{b2Req4}
\end{gather}
This confirms that the restricted metric of the $B_2$ gravity can also be described by a spin-one Abelian gauge potential. 

Notice that in the above $B_2$ gravity we have chosen 
the lightlike isometry to be $\vj$ (and $\vtj$). But obviously, we could also have chosen the Abelian direction to be $\vk$ (and $\vtk$). In this case 
the $B_2$ gravity can be expressed by
\begin{gather}
\cL =\dfrac{e}{16\pi G_N} \Big[\vg_\mn \cdot \vR_\mn 
+\vlam_\mu \cdot D_\mu \vk \Big]   \nn\\
=\dfrac{e}{16\pi G_N} \Big[\hvg_\mn \cdot \hvR_\mn 
+\vlam_\mu \cdot \vZ_\mu \Big],  
\label{b2rg3}
\end{gather}
where now $\hvR_\mn$ and $\vZ_\mu$ here are fixed by 
the constraint $D_\mu \vk=0$.    

Clearly both (\ref{a2Req4}) and (\ref{b2Req4}) imply that the dynamical field (the restricted metric) of the restricted gravity is described by a massless spin-one field. But this is the only dynamical 
degrees that we have, so that this must be identified as the graviton. This strongly implies that 
the graviton, just like the photon, can indeed be described by a massless spin-one field. 

So far we have discussed the Lagrangian formalism 
of the restricted gravity in the absence of outside 
gravitational source, with no coupling to the energy-momentum tensor. However, we can easily extend the formalism 
in the presence of the energy-momentum tensor given by
\begin{gather}
T_{\mn} = \frac{2}{\sqrt{-g}} 
\frac{\delta \mathcal{S}_{m}}{\delta g^{\mn}} \ ,
\end{gather}
where $\mathcal{S}_{m}$ is the action of matter fields. 
In the presence of the matter fields, the first equation 
of the $A_2$ gravity in \eqref{a2fulleqvac} is 
modified to
\begin{gather}
G_\mn (\cD^\nu \bA^\rho-\cD^\rho \bA^\nu)
-\tG_\mn (\cD^\nu B^\rho-\cD^\rho B^\nu) \nn\\
+\vG_\mn \cdot (\hD^\nu \vZ^\rho-\hD^\rho \vZ^\nu)
= \kappa \bigg( \tensor{T}{_\mu ^\rho}  
- \frac{1}{2}\tensor{\delta}{_\mu ^\rho} T \bigg), 
\end{gather}
where $\kappa = 8\pi G$ is the gravitational coupling, 
and $T$ is the trace of the energy-momentum tensor. 

Similarly, for the $B_2$ gravity the first equation 
of \eqref{b2fulleqvac} is modified to
\begin{gather}
\cJ_\mn (\cD^\nu K^\rho-\cD^\rho K^\nu)
-\ctJ_\mn (\cD^\nu \tK^\rho-\cD^\rho \tK^\nu) \nn\\
+\cK_\mn (\cD^\nu J^\rho-\cD^\rho J^\nu)
-\ctK_\mn (\cD^\nu \tJ^\rho-\cD^\rho \tJ^\nu) \nn\\
+\vG_\mn \cdot \vZ^{\nu\rho} 
=\kappa \bigg( \tensor{T}{_\mu ^\rho}  
- \frac{1}{2}\tensor{\delta}{_\mu ^\rho} T \bigg).
\end{gather}
Notice that the remaining equations in \eqref{a2fulleqvac} 
and \eqref{b2fulleqvac} do not change since they are 
the metric-compatibility conditions.

\section{Spacetime geometry of restricted gravity with trivial topology}

One may wonder what geometrical properties of 
the spacetime the restricted gravity can have. To 
discuss this question we have to remember that 
the restricted gravity has topological part coming from the non-trivial topology of the Abelian direction. For instance, the restricted $A_2$ 
curvature (\ref{a2rct}) contains the topological 
part $H_\mn$ which comes from the non-trivial 
topology of $\vl$. Similarly, the restricted $B_2$ curvature (\ref{b2rct}) contains the topological 
part $H_\mn$ generated by the non-trivial topology 
of $\vj$. This allows the restricted gravity to inherit all topological properties of Einstein's 
theory \cite{cqg12,cqg13,gc15}.

In the following, however, we consider the restricted gravity with trivial topology for simplicity. 
This simplifies the theory and allows us to discuss the geometicall properties of the restricted gravity more easily. With this assumption we show that all null tetrads of the spacetime described by $A_2$ gravity are recurrent, which means that the spacetime of $A_2$ gravity with trivial topology is a direct product of a two-dimensional Lorentzian manifold and 
a two-dimensional Riemannian manifold. Moreover, we show that the spacetime of $B_2$ gravity with trivial topology admits a covariantly constant null vector field, which means that the spacetime described by 
the $B_2$ gravity belongs to the pp-wave spacetimes. 

So in the following we assume the orthonormal basis vectors $\hn_{i}$ to be topologically trivial, that is, independent of the spacetime coordinates and 
have the form 
\begin{gather}
(\hn_i)_j = \delta_{ij},~~~(i,j=1,2,3).
\label{orthobasis}
\end{gather}
This guarantees that the restricted gravity 
has no non-trivial topology. 

\subsection{$A_2$ Gravity with Trivial Topology}

In $A_2$ gravity, the nonvanishing components
of the spin connection are given by
\begin{gather}
\Gm_\mu^{12} = - \Gm_\mu^{21} = A_\mu, 
~~~~\Gm_\mu^{03} = -\Gm_\mu^{30} = B_\mu, 
\end{gather} 
and the covariant derivatives of the tetrads are written in terms of the gauge potentials as
\begin{gather}
\nabla_\mu (e^{\hat{0}})_\nu 
= B_\mu (e^{\hat{3}})_\nu, 
~~~~\nabla_\mu (e^{\hat{1}})_\nu 
= A_\mu (e^{\hat{2}})_\nu, \nn \\
\nabla_\mu (e^{\hat{2}})_\nu 
= A_\mu (e^{\hat{1}})_\nu,
~~~~\nabla_\mu (e^{\hat{3}})_\nu 
= B_\mu (e^{\hat{0}})_\nu.
\label{a2cdtetrads}
\end{gather}
So, with the four null tetrads given by
\begin{gather}
k_\mu = \frac{1}{\sqrt{2}} 
(e^{\hat{0}}+e^{\hat{3}})_\mu,
~~~l_\mu = \frac{1}{\sqrt{2}} 
(e^{\hat{0}}-e^{\hat{3}})_\mu, \nn \\
m_\mu = \frac{1}{\sqrt{2}}
(e^{\hat{1}}+ i e^{\hat{2}})_\mu, 
~~~\bar{m}_\mu = \frac{1}{\sqrt{2}}
(e^{\hat{1}}- i e^{\hat{2}})_\mu,
\end{gather}
we can express (\ref{a2cdtetrads}) as
\begin{gather}
\nabla_\mu k_\nu = B_\mu k_\nu, 
~~~\nabla_\mu l_\nu = -B_\mu l_\nu, \nn \\
\nabla_\mu m_\nu = -i A_\mu m_\nu, 
~~~\nabla_\mu \bar m_\nu = i A_\mu \bar m_\nu.
\label{a2cdntetrads}
\end{gather}
This tells that the $A_2$ gravity admits four 
recurrent null vector fields $k_\mu$, $l_\mu$, 
$m_\mu$, and $\bar m_\mu$. Thus, there exist four 
real functions $f$, $\tilde f$, $u$, and $v$, and 
two complex functions $h$ and $\zeta$ such that \cite{ehlers},
\begin{gather}
k_\mu dx^\mu = f du, 
~~~l_\mu dx^\mu = \tilde f dv, \nn \\
m_\mu dx^\mu = h~d\zeta, 
~~~\bar m_\mu dx^\mu = \bar h~d\bar \zeta.
\end{gather}
With this the metric can be written as
\begin{gather}
ds^{2} = -f \tilde f~du dv 
+ 2 h\bar h~ d\zeta d\bar \zeta.
\label{a2metric1}
\end{gather}
From (\ref{a2cdntetrads}) we find that all Newman-Penrose spin connection coefficients vanish except $\alpha$, $\beta$, $\gamma$, and $\epsilon$. 
In particular, $\pi=\tau=0$ and $\mu = \rho=0$ imply
\begin{gather}
(f\tilde f)_{,\zeta} =(f\tilde f)_{,\bar \zeta}=0, \nn \\
(h\bar h)_{,u} = (h\bar h)_{,v} = 0,
\end{gather}
respectively. This tells us that the spacetime that represents the $A_2$ gravity with trivial topology 
is a direct product of a two-dimensional Lorentzian manifold $M_2$ and a two-dimensional Riemannian manifold $\Sigma_2$. 

Let us introduce a new coordinates $(t,x,y,z)$ by
\begin{gather}
u=\frac{1}{\sqrt{2}} (t-z), 
~~~v=\frac{1}{\sqrt{2}} (t+z),  \nn\\
\zeta = \frac{1}{\sqrt{2}} (x+iy),
~~~\bar \zeta = \frac{1}{\sqrt{2}} (x-iy).
\end{gather}
In this coordinates the metric (\ref{a2metric1}) acquires the form
\begin{gather}
ds^{2} = e^{2M(t,z)} (-dt^{2} + dz^{2}) 
+ e^{2N(x,y)} ( dx^{2} + dy^{2} ).
\label{a2metric}
\end{gather}
for some functions $M$ and $N$. Moerover, with 
the tetrad defined by
\begin{gather}
e_{\hat{0}} = e^{-M}\partial_{t},
~~~e_{\hat{1}} = e^{-N}\partial_{x}, \nn\\
e_{\hat{2}} = e^{-N}\partial_{y}, 
~~~e_{\hat{3}} = e^{-M}\partial_{z}, 
\label{a2tetrads}
\end{gather}
the restricted metric $G^{\mn}$ and $\tG^{\mn}$ 
are given by
\begin{gather}
G^\mn = e^{-2N} \big[dx^\mu dy^\nu 
-dx^\nu dy^\mu \big], \nn\\
\tG^{\mn} = e^{-2M} \big[ (dt)^{\mu} (dz)^{\nu}
- (dt)^{\nu} (dz)^{\mu} \big],
\label{a2rm}
\end{gather}
for which the metricity condition 
\begin{gather}
\nabla_\mu G^\mn =\nabla_\mu \tG^\mn = 0
\end{gather}
automatically hold. From this we have
\begin{gather}
G_\mn= \pd_\mu G_\nu -\pd_\nu G_\mu,
~~~G_\mu=\Big(\Int e^{2N} dx \Big) \pd_\mu y,   \nn\\
\tilde G_\mn= \pd_\mu \tilde G_\nu 
-\pd_\nu \tilde G_\mu,
~~~\tilde G_\mu=\Big(\Int e^{2M} dt \Big) \pd_\mu z,  	
\end{gather}
where $G_\mu$ and $\tilde G_\mu$ are the spin-one potentials which describe the restricted metric $G_\mn$ and $\tilde G_\mn$. 

Moreover, the $A_2$ restricted connection and 
the corresponding curvature tensor are given by
\begin{gather}
A_{\mu} = (\partial_{y} N) \partial_{\mu}x
- (\partial_{x} N)\partial_{\mu} y, \nn\\
B_{\mu} = -(\partial_{z} M)\, \partial_{\mu} t
- (\partial_{t} M)\, \partial_{\mu}z,  \nn\\
A_\mn = -\big(\pd_x^2 N +\pd_y^2 N \big) 
(\pd_\mu x \pd_\nu y -\pd_\nu x \pd_\mu y), \nn\\
B_\mn = \big(-\pd_t^2 M +\pd_z^2 M \big)
(\pd_\mu t \pd_\nu z -\pd_\nu t \pd_\mu z).
\label{a2fs}
\end{gather}
So, from (\ref{a2rm}) and (\ref{a2fs}) we have
\begin{gather}
A_{\mu \sigma} \tensor {G}{^\sigma_\nu}
- B_{\mu \sigma} \tensor {\tG}{^\sigma_\nu} \nn\\
= \big(-\pd_t^2 M + \pd_z^2 M)
(-\pd_\mu t \pd_\nu t +\pd_\mu z \pd_\nu z \big) \nn\\
+ \big(\pd_x^2 N + \pd_y^2 N \big) 
(\pd_\mu x \pd_\nu x + \pd_\mu y \pd_\nu y).
\label{a2Ricci}
\end{gather}
With this we can solve the equation of motion 
of the $A_2$ gravity (\ref{a2Req0}).

\subsection{$B_2$ Gravity with Trivial Topology}

The nonvanishing components of the restricted connection of $B_2$ gravity are 
\begin{gather}
\Gm_\mu^{01} = - \Gm_\mu^{10} = -\Gm_\mu^{31} 
= \Gm_\mu^{13} = \frac{1}{\sqrt 2} \tGm_\mu, \nn\\
\Gm_\mu^{02} = - \Gm_\mu^{20} = \Gm_\mu^{23} 
= -\Gm_\mu^{32} = \frac{1}{\sqrt 2} \Gm_\mu. 
\end{gather}
It follows from these that the covariant 
derivatives of the tetrads are expressed as
\begin{gather}
\nabla_\mu (e^{\hat 0})_\nu 
=-\nabla_\mu (e^{\hat 3})_\nu 
= \frac{1}{\sqrt 2} \big[\tGm_\mu (e^{\hat 1})_\nu
+ \Gm_\mu (e^{\hat 2})_\nu \big], \nn\\
\nabla_\mu (e^{\hat 1})_\nu =\frac{1}{\sqrt 2} 
\tGm_\mu (e^{\hat 0}+e^{\hat 3})_\nu, \nn\\
\nabla_\mu (e^{\hat 2})_\nu =\frac{1}{\sqrt 2} 
\Gm_\mu (e^{\hat 0}+e^{\hat 3})_\nu, 
\label{b2cdtetrads}
\end{gather}
or equivalently,
\begin{gather}
\nabla_\mu k_\nu = 0, 
~~~\nabla_\mu l_\nu = \bar \eta_\mu m_\nu
+ \eta_\mu \bar m_\mu,   \nn \\
\nabla_\mu m_\nu = \eta_\mu k_\nu, 
~~~\nabla_\mu \bar m_\nu = \bar \eta_\mu k_\nu,
\label{b2cdntetrads}
\end{gather}
where $\eta_{\mu} = \tGm_{\mu} + i \Gm_{\mu}$. 
Since $k_{\mu}$ is covariantly constant, the spacetime of $B_2$ gravity belongs to the pp-wave spacetimes 
with the metric given by \cite{ehlers},
\begin{gather}
ds^{2} =-2 du \big[H(u,x,y) du + dv \big] 
+ dx^2 + dy^2,
\label{b2metric}
\end{gather}
for a function $H$. 

Let us choose the tetrad defined by
\begin{gather}
e_{\hat 0} = -\frac{1}{\sqrt 2} \pd_u 
- \frac{(1-H)}{\sqrt 2} \pd_v, \nn \\
e_{\hat 3} = -\frac{1}{\sqrt 2} \pd_u 
+ \frac{(1+H)}{\sqrt 2} \pd_v, \nn \\
e_{\hat 1} = \pd_x,~~~e_{\hat 2} =\pd_y.
\end{gather}
With this we have the restricted metric $\cJ_\mn$ 
and $\ctJ_\mn$ 
\begin{gather}
\cJ_\mn =\pd_\mu u \pd_\nu y -\pd_\nu u \pd_\nu y, \nn\\
\ctJ_\mn =\pd_\mu u \pd_\nu x -\pd_\nu u \pd_\nu x. 
\label{b2rm}
\end{gather} 
This trivially satisfies the metricity condition
\begin{gather}
\nabla_\mu \cJ^\mn = \nabla_\mu \ctJ^\mn = 0,
\end{gather}
so that we have
\begin{gather}	
\cJ_\mn= \pd_\mu \cJ_\nu -\pd_\nu \cJ_\mu,
~~~\cJ_\mu= u \pd_\mu y,   \nn\\
\tilde \cJ_\mn= \pd_\mu \tilde \cJ_\nu 
-\pd_\nu \tilde \cJ_\mu,
~~~\tilde \cJ_\mu=-u \pd_\mu x,  	
\end{gather}
where $\cJ_\mu$ and $\tilde \cJ_\mu$ are the spin-one potentials of the restricted metric $\cJ_\mn$ and $\tilde \cJ_\mn$.

The $B_2$ gravitational connection and curvature tensor associated with the tetrad are given by 
\begin{gather}
\Gm_\mu =-(\pd_y H)~\pd_\mu u, 
~~~\tGm_\mu = -(\pd_x H)~\pd_\mu u,   \nn\\
K_\mn =(\pd_x \pd_y H) (\pd_\mu u~\pd_\nu x 
-\pd_\nu u~\pd_\mu x)  \nn\\
+(\pd_y^2 H) (\pd_\mu u~\pd_\nu y 
-\pd_\nu u~\pd_\mu y), \nn\\
\tK_\mn =(\pd_x^2 H) (\pd_\mu u~\pd_\nu x 
-\pd_{\nu}u~\pd_\mu x) \nn\\
+(\pd_x \pd_y H) (\pd_\mu u~\pd_\nu y 
-\pd_\nu u~\pd_\mu y). 
\label{b2fs}
\end{gather}
And from (\ref{b2rm})and  (\ref{b2fs}) we have
\begin{gather}
K_{\mu \sigma} \tensor{\cJ}{^{\sigma}_{\nu}} 
- \tK_{\mu \sigma} \tensor{\ctJ}{^{\sigma}_{\nu}}
= -(\Delta H)~(\pd_\mu u~\pd_\nu u),
\label{nullricci}
\end{gather} 
where $\Delta = \pd_x^2 + \pd_y^2$. This tells that 
the matter fields described by the $B_2$ gravity 
are pure radiations (null dusts). With this we can 
solve the restricted $B_2$ equation of motion 
(\ref{b2Req0}).

\section{Examples of $A_2$ gravity}

In this section we present several explicit examples of the spacetimes of $A_2$ gravity. As was discussed, the $A_2$ gravity with trivial topology describes 
the spacetimes that are direct products of 
the Lorentzian and Riemaniann 2-surfaces, 
$M_2 \times \Sigma_{2}$. There are various solutions 
of such spacetimes, for instance, the Nariai and 
the anti-Nariai spacetimes are the direct products $dS_{2}\times S^{2}$ and $AdS_{2} \times H^2$ \cite{nariai,anariai}. They are the solutions to 
the vacuum Einstein's equations with the positive 
and negative cosmological constants. 

The Bertotti-Robinson solution describes 
an electrovacuum without cosmological constant, 
and the corresponding spacetime is a direct product 
of $AdS_2 \times S^2$ \cite{bertotti,robinson}. 
In the following subsections, we will construct 
two exact solutions to the $A_2$ gravity with $\Sigma_2$ identified with $E_{2}$, the Euclidean two-plane, and $S^{2}$. These solutions will turn 
out to be the cosmic string and the Bertotti-Robinson spacetimes, respectively.

\subsection{Gravitational Cosmic String}

As the simplest example of the $A_2$ gravity, 
we discuss a static and axisymmetric spacetime. 
Consider the coordinate transformation 
\begin{gather}
x= h(\rho) \cos (\phi),
~~~~y= h(\rho) \sin (\phi), 
\label{a2cc1}
\end{gather}
where $h=h(\rho)$ is an arbitrary function. 
Under this transformation the metric (\ref{a2metric}) transforms to
\begin{gather}
ds^2 = e^{2M} (-dt^2 + dz^2) + e^{2 \tilde N} 
(d\rho^2 + \tilde h^2 d\phi^2),  \nn\\
\tilde h = h \bigg(\frac{dh}{d\rho} \bigg)^{-1},
~~~\tilde N =N + \ln \bigg|\frac{dh}{d\rho}\bigg|.
\end{gather}
Notice that $dh/d\rho$ should not vanish to make 
the Jacobian of the coordinate transform (\ref{a2cc1}) nonzero. Suppose we have two commuting Killing 
vectors $\pd_t$ and $\pd_\phi$ so that $M$ solely depends on $z$, and $\tilde N$ and $f$ depend
only on $\rho$. In this case we can introduce 
a proper radius $R$ on $\Sigma_{2}$ and a new 
function $a(\rho)$ defined by
\begin{gather}
R(\rho) = \int e^{\tilde N} \tilde h~d\rho, 
~~~a(R) = e^{\tilde N \rho(R)} \tilde h~\rho(R),
\end{gather}
where $R(\rho)$ is assumed to be invertible. 

With this the metric becomes
\begin{gather}
ds^{2} = e^{2M(z)} (-dt^2 + dz^2) + dR^2 
+ a^2 (R) d\phi^{2},
\end{gather}
and the restricted metric $G_\mn$ and $\tilde G_\mn$ are given by 
\begin{gather}
G_\mn= a(R) (\pd_\mu R \pd_\nu \phi 
-\pd_\nu R \pd_\mu \phi),   \nn\\
\tilde G_\mn= -e^{2M} (\pd_\mu t \pd_\nu z
-\pd_\nu t \pd_\mu z),   \nn\\
G_\mu =\Big(\int a(R) dR \Big) \pd_\mu \phi,  \nn\\
~~~\tilde G_\mu=\Big(\int e^{2M} dt \Big) \pd_\mu z,  	
\end{gather}
where $G_\mu$ and $\tilde G_\mu$ are the spin-one potentials of the restricted metric. Moreover, 
the restricted connection and curvature tensor 
are given by
\begin{gather}
A_\mu = -\bigg(\frac{da}{dR}\bigg) \pd_\mu \phi, 
~~~B_\mu = -\bigg(\frac{dM}{dz} \bigg) \pd_\mu t, \nn\\
A_\mn =-\bigg(\frac{d^2 a}{dR^2} \bigg)
(\pd_\mu R \pd_\nu \phi -\pd_\nu R \pd_\mu \phi), \nn\\
B_\mn =\bigg(\frac{d^{2}M}{dz^2} \bigg) 
~(\pd_\mu t \pd_\nu z -\pd_\nu t \pd_\mu z).
\end{gather}
With this the equation of motion (\ref{a2Ricci}) becomes
\begin{gather}
A_{\mu \sigma} \tensor{G}{^{\sigma}_{\nu}}
- B_{\mu \sigma} \tensor{\tG}{^{\sigma}_{\nu}} 
= \bigg(\frac{d^2 M}{dz^2} \bigg)
(-\pd_\mu t \pd_\nu t +\pd_\mu z \pd_\mu z \big) \nn\\
+\bigg(\frac{1}{a} \frac{d^2 a}{dR^2} \bigg)
(\pd_\mu R \pd_\nu R +a^2 \pd_\mu \phi \pd_\nu \phi).
\end{gather}
Now, for the vacuum spacetime, $M(z)$ and $a(R)$ should be linear, so that we have
\begin{gather}
M(z) = m_0 z + m_1,~~~a(R) = a_0 R + a_1,
\end{gather} 
for some constant $m_0$, $m_1$, $a_0$, and $a_1$. 
To avoid the singularity along the $z$-axis, 
we impose $a_1=0$. Moreover, we can eliminate 
the conformal factor $e^{2M}$ from the metric through the coordinate transformation 
\begin{gather}
T= e^{m_0 z + m_1} \sinh(t), 
~~~Z= e^{m_0 z + m_1} \cosh(t).
\end{gather}
The resulting metric becomes
\begin{gather}
ds^2 = -dT^2 + dZ^2 + dR^2 + a_0^2 R^2 d \phi^2. 
\label{cstring}
\end{gather}
where the constant $a_0$ measures the angle defect 
of the conical singularity along the $z$-axis. 
As is well known, this singularity can be removed 
introducing a tube along the $z$-axis with uniform mass distribution given by \cite{hiscock, gott}
\begin{gather}
T_\mu^\nu= \left\{\begin{array}{ll}
-\epsilon(\pd_\mu t ~\pd^\nu t +\pd_\mu z~\pd^\nu z), 
& \textrm{~~($R < R_0$)} \\ 
0, & \textrm{~~($R > R_0$)} 
\end{array} \right.
\end{gather} 
where $\epsilon$ is the mass density and $R_0$ is 
the radius of the tube. 

With this the equations of motion for $R < R_0$ reduces to a single equation
\begin{gather}
a''+8\pi \epsilon a = 0.
\label{a2eq2}
\end{gather}
From this we have
\begin{gather}
f=A \sin{\bigg(\frac{R}{R_{\ast}}\bigg)} 
+B \cos{\bigg(\frac{R}{R_{\ast}}\bigg)},
\end{gather}
where $R_\ast = 1/\sqrt{8 \pi \epsilon}$. Moreover, 
assuming that the metric is flat with no conical 
sigularity near the axis, we have
\bea
A=R_\ast, ~~~B=0,
\eea
so that
\bea
ds^2=-dT^2 + dZ^2 +dR^2 + R_\ast^2 
\sin^2{(\frac{R}{R_\ast})}~d\phi^2.
\eea
For the exterior we must have (\ref{cstring}). 
Connecting the interior and exterior solutions 
smoothly, we have
\begin{gather}
ds^2=-dT^2 + dZ^2 +dR^2 +R^2 \cos^2 {\bigg(\frac{R_0}
{R_\ast}\bigg)}d\phi^2, 
\end{gather}
which avoids the surface-stress energy tensor at $R=R_0$. This is the gravitational cosmic string, which shows that the $A_2$ gravity can describe 
the gravitationsl cosmic string.

\subsection{Bertotti-Robinson Spacetime}

Next, consider the static and spherically symmetric system of $A_2$ gravity. Let us assume that 
the transverse two-surface is a two-sphere ($\Sigma_2=S^2$), and introduce the new coordinates $(t,r,\theta,\phi)$ given by
\begin{gather}
u=\frac{1}{\sqrt 2} (t-r),
~~~v=\frac{1}{\sqrt 2} (t+r),  \nn \\
x =\cot \frac{\theta}2~\cos \phi, 
~~~y =\cot \frac{\theta}2~\sin \phi .
\end{gather}
In this cordinate the metric (\ref{a2metric}) becomes
\begin{gather}
ds^2 = e^{2P(r)} (-dt^2 + dr^2) \nn\\
+ e^{2Q(\theta)} ( d\theta^2 + \sin^2\theta d\phi^2).
\label{brmetric}
\end{gather}
Choosing the tetrad defined by
\begin{gather}
e_{\hat 0} = e^{-P} \pd_t,
~~~e_{\hat 1} = e^{-Q} \pd_\theta, \nn \\
e_{\hat 2} = e^{-Q} \csc \theta~\pd_\phi, 
~~~e_{\hat 3} = e^{-P} \pd_r, 
\label{brtetrads}
\end{gather}
we have the restricted metric $G_\mn$ and $\tG_\mn$ expressed by
\begin{gather}
G_\mn = e^{2Q} \sin \theta~\big(\pd_\mu \theta 
\pd_\nu \phi -\pd_\nu \theta \pd_\mu \phi \big), \nn\\
\tG^\mn = e^{2P} \big(\pd_\mu t \pd_\nu r 
-\pd_\nu t \pd_\mu r \big),  \nn\\
G_\mu=\Big(\int e^{2Q} \sin \theta d\theta \Big) \pd_\mu \phi,   \nn\\
\tilde G_\mu=\Big(\int e^{2P} dr \Big) \pd_\mu t,
\label{brrm}
\end{gather}
where $G_\mu$ and $\tilde G_\mu$ are the spin-one potentials of the restricted metric. 

Moreover, we have the restricted connection and 
the corresponding curvature tensor given by
\begin{gather}
A_\mu = -\bigg(\cos \theta +\sin\theta \frac{dQ}{d\theta}\bigg)~\pd_\mu \phi,
~~~B_\mu =-\bigg(\frac{dP}{dr} \bigg)~\pd_\mu t, \nn\\
A_\mn =-\sin \theta \bigg(\frac{d^2 G}{d\theta^2}
+ \cot \theta~\frac{dG}{d\theta}- 1 \bigg)  \nn\\
\times (\pd_\mu \theta \pd_\nu \phi 
-\pd_\nu \theta \pd_\mu \phi), \nn\\
B_\mn =\bigg(\frac{d^2P}{dr^2} \bigg) 
~(\pd_\mu t \pd_\nu r -\pd_\nu t \pd_\mu r).
\label{brfs}
\end{gather}
So the field equation becomes
\begin{gather}
A_{\mu \sigma} \tensor{G}{^{\sigma}_{\nu}} 
- B_{\mu \sigma} \tensor{\tG}{^{\sigma}_{\nu}}
=\bigg(\frac{d^2P}{dr^2} \bigg)(-\pd_\mu t \pd_\nu t
+ \pd_\mu r \pd_\nu r) \nn\\
+\bigg(\frac{d^2 Q}{d\theta^2}
+ \cot \theta~\frac{dQ}{d\theta}- 1 \bigg)  
(\pd_\mu \theta \pd_\nu \theta   \nn\\
+ \sin^2 \theta~\pd_\mu \phi \pd_\nu \phi). 
\end{gather}
Now, let us consider the static and spherically symmetric electromagnetic field strength in the form of
\begin{gather}
F_\mn = f(r) (\pd_\mu t \pd_\nu r 
-\pd_\nu t \pd_\mu r),
\end{gather}
whose energy-momentum tensor is given by
\begin{gather}
T_\mn =\frac{f^2}{2}e^{-2P} \big(\pd_\mu t \pd_\nu t 
- \pd_\mu r \pd_\nu r \big) \nn\\
+ \frac{f^{2}}{2}e^{2Q-4P} \big(\pd_\mu \theta
\pd_\nu \theta + \sin^2 \theta
~\pd_\mu \phi \pd_\nu \phi \big).
\end{gather}
With this the equations of motion reduce to 
\begin{gather}
\frac{d^2 P}{dr^2} - 4 \pi f^2 e^{-2P} =0, \nn\\
\frac{d^2 Q}{d \theta^2}
+ \cot \theta ~\frac{dQ}{d\theta}
+ 4 \pi f^2 e^{2Q-4P} - 1 =0.
\end{gather}
The last equation tells that $4 \pi f^2 e^{-4P} 
\equiv C_{1}$ should be a positive constant, so that 
the equation reduces to
\begin{gather}
\frac{d^2 P}{dr^2} - C_{1} e^{2P} =0, \nn\\
\frac{d^2 Q}{d\theta^2}
+\cot \theta~\frac{dQ}{d\theta} +C_{1}e^{2Q} -1 =0.
\end{gather}
From this we have
\begin{gather}
P(r) = -\frac12 \ln \big[ \sqrt{C_{1}}( r+C_{2})^2 \big],
\end{gather} 
where $C_{2}$ is a constant that can be eliminated 
by a suitable translation of $r$. 

It is convenient to introduce a new parameter 
$q=\pm C_1 \sqrt{4\pi}$ so that electromagnetic 
field strength becomes
\begin{gather}
F_\mn = \frac{q}{4\pi r^{2}}~(\pd_\mu t \pd_\nu r 
-\pd_\nu t \pd_\mu r). 
\end{gather}
It can be readily checked that the regular solution 
to the equation for $Q$ is 
\begin{gather}
e^{2Q} = \frac{q^{2}}{4\pi}. 
\end{gather}
Finally, the metric of this spacetime can be written in a conformally flat form as
\begin{gather}
ds^2 = \frac{q^2}{4\pi r^2} \big( -dt^2 + dr^2 \nn\\
+ r^2 d\theta^2 + r^2 \sin^2\theta~d\phi^2 \big),
\end{gather}
and the restricted connection and the curvature tensor (\ref{brfs}) are simplified to
\begin{gather}
A_\mu = -\cos \theta~\pd_\mu \phi, 
~~~B_\mu =\frac{1}{r} \pd_\mu t, \nn\\
A_\mn = \sin \theta~(\pd_\mu \theta \pd_\nu \phi 
-\pd_\nu \theta \pd_\mu \phi), \nn\\
B_\mn =\frac{1}{r^2} (\pd_\mu t \pd_\nu r 
-\pd_\nu t \pd_\mu r).
\end{gather}
This solution, discovered independently by Bertotti and Robinson \cite{bertotti,robinson}, describes a universe with homogeneous electromagnetic fields. This confirms that 
the $A_2$ gravity can describe 
the Bertotti-Robinson spacetime. 

\section{Examples of $B_2$ Gravity}

In this section we present several explicit examples 
of the $B_2$ gravity. As a counterpart of the cosmic string in the $A_2$ gravity, we construct an axisymmetric pp-wave solution whose singularity 
along the symmetry axis can be removed introducing 
a uniform beam of massless fields such as electromagnetic radiation fields. This solution 
turns out to be the Bonnor's beam \cite{bonnor}. 

We also construct a conformally flat spacetime that belongs to the $B_2$ gravity by assuming that the wavefront is planar in the sense that the curvature on each wavefront is homogeneous. This spacetime describes the Cahen-Leroy spacetime \cite{cahen}. 

\subsection{Axisymmetric pp-wave}

Consider a stationary and axisymmetric 
$B_2$ gravity. Choose the following coordinates
\begin{gather}
x= \rho \cos \phi,~~~y=\rho \sin \phi,
\end{gather}
and the metric
\begin{gather}
ds^2 =- 2du \big[H(u,\rho,\phi)du +dv \big] 
+d\rho^2 + \rho^2 d\phi^2. 
\end{gather}
With this we have the restricted metric $\cJ_\mn$ 
and $\ctJ_\mn$ 
\begin{gather}
\cJ_\mn =\sin \phi (\pd_\mu u \pd_\nu \rho 
-\pd_\nu u \pd_\mu \rho)  \nn\\
+ \rho \cos \phi 
(\pd_\mu u \pd_\nu \phi-\pd_\nu u \pd_\mu \phi), \nn\\
\ctJ_\mn =\cos \phi (\pd_\mu u \pd_\nu \rho 
-\pd_\nu u \pd_\mu \rho) \nn\\
+ \rho \sin \phi 
(\pd_\mu u \pd_\nu \phi-\pd_\nu u \pd_\mu \phi),  \nn\\
\cJ_\mu =u\sin \phi \pd_\mu \rho
+u \rho \cos \phi \pd_\mu \phi,  \nn\\
\tilde \cJ_\mu =-u \cos \phi \pd_\mu \rho
+u \rho \sin \phi \pd_\mu \phi,  
\label{b2rm1}
\end{gather} 
where $\cJ_\mu$ (and $\tilde \cJ_\mu$) is the spin-one potential which describes the restricted metric $\cJ_\mn$ (and $\tilde \cJ_\mn$). Moreover, 
the corresponding $B_2$ connection and curvature 
tensor are given by
\begin{gather}
\Gm_\mu =-\frac{1}{\sqrt 2} \dot H \sin \phi
(\pd_\mu t -\pd_\mu z), \nn\\
\tGm_\mu =-\frac{1}{\sqrt 2} \dot H \cos \phi
(\pd_\mu t -\pd_\mu z), \nn\\
K_\mn =\frac{1}{\sqrt 2} \ddot H \sin \phi
~(\pd_\mu t \pd_\nu \rho -\pd_\nu t \pd_\mu \rho \nn\\
+ \pd_\mu \rho \pd_\nu z -\pd_\nu \rho \pd_\mu z), \nn\\
\tK_\mn =\frac{1}{\sqrt 2} \ddot H \cos \phi 
~(\pd_\mu t \pd_\nu \rho -\pd_\nu t \pd_\mu \rho \nn\\
+\pd_\mu \rho \pd_\nu z -\pd_\nu \rho \pd_\mu z),
\end{gather}
where the dot denotes the derivative with respect to $\rho$. 

Now, assume that the spacetime is axisymmetric  
along the $z$-axis and stationary. In this case  
$H$ depends only on $\rho$, and the equation of motion (\ref{nullricci}) becomes
\begin{gather}
K_{\mu \sigma} \tensor{\cJ}{^{\sigma}_{\nu}} 
-\tK_{\mu \sigma} \tensor{\ctJ}{^{\sigma}_{\nu}} \nn\\
=-\big(\ddot H+ \frac{1}{\rho} \dot H \big)
~(\pd_\mu u~\pd_\nu u). 
\end{gather} 
The solution to the vacuum field equations is 
found to be
\begin{gather}
H_{\textrm{ext}}(\rho) = h_1 \ln (\rho) + h_2,
\end{gather}
for some constants $h_1$ and $h_2$. As in the case 
of the cosmic string, the singularity along the $z$ axis can be removed introducing a uniform distribution of a radiation propagating along the $z$-axis,
\begin{gather}
T_\mn = \left\{\begin{array}{ll}
\epsilon_2 (\pd_\mu u~\pd_\nu u),
& \textrm{~~($\rho < \rho_0$)} \\ 0, 
& \textrm{~~($\rho > \rho_0$)} 
\end{array} \right.
\end{gather} 
where $\rho_0$ and $\epsilon_2$ are the cross-sectional radius and the energy density of 
the beam. 

The interior solution is given by
\begin{gather}
H_{\textrm{int}}(\rho) = 2\pi \epsilon_2 \rho^2,
\end{gather}
where the possible integration constants were neglected to avoid the singularity along the $z$-axis. With the differentibility and the continuity condition of $H$ on the boundary $\rho=\rho_0$
\begin{gather}
h_1 = 4\pi \epsilon_2 \rho_0^2, 
~~~h_2 = 2 \pi \epsilon_2 \rho_0^2 (1-\ln \rho_0^2),
\end{gather}
we have the exterior solution $H_{\textrm{ext}}$ 
\begin{gather}
H_{\textrm{ext}}(\rho) = 2 \pi \epsilon_2 \rho_0^2  \big[\ln (\rho^2/\rho_0^2) +1 \big]. 
\end{gather}
This confirms that the $B_2$ gravity has 
the axis-symmetric plane-fronted wave with parallel propagation (i.e., the pp-wave) as 
a solution \cite{ehlers,bonnor}.

\subsection{Conformally Flat Wave with Planar Wavefront}

As we have seen in the previous section, the $A_2$ gravity contains the Bertotti-Robinson spacetime 
that is the only conformally flat solution to 
the Einstein-Maxwell equations associated 
with non-null electromagnetic fields. In this subsection we construct a counterpart of 
the Bertotti-Robinson spacetime which is described 
by $B_2$ gravity associated with the null electromagnetic field.

Consider the $B_2$ gravity with the metric 
(\ref{b2metric}). It follows from (\ref{nullricci}) that the energy-momentum tensor takes 
the form of
\begin{gather}
T_\mn = \frac{1}{2}\bar \Phi \Phi~\pd_\mu u \pd_\nu u,
\label{nullemt}
\end{gather}
for some complex function $\Phi=\Phi(u)$. Any electromagnetic field which has the above energy-momentum tensor can be cast into 
the following form,
\begin{gather}
\tilde F_\mn =\Phi(u) (k_\mu m_\nu-k_\nu m_\mu),
\end{gather}
where $\tilde F$ is the dual of $F$. With this 
the equation of motion becomes
\begin{gather}
(\pd_x^2 +\pd_y^2) H = 4\pi \bar \Phi \Phi.
\label{b2eqcf}
\end{gather}
For simplicity, let us assume that the pp-wave is planar, that is, the curvature is homogeneous on 
each wavefront. In our coordinates, this is equivalent to the statement that all third derivatives of $H$ with respect to $x$ and $y$ vanish \cite{roche}. 
So the metric function $H$ is at most quadratic in $x$ and $y$. Moreover, the terms of $H$ that are linear in $x$ and $y$ can be eliminated by a suitable coordinate transformation \cite{roche, mclenaghan}. Thus, $H$ can be written as  
\begin{gather}
H(u,x,y) =a(u)(x^2 + y^2) + b(u) (x^2-y^2) \nn\\
+ c(u)~xy,
\end{gather}   
with $u$-dependent coefficients. 

It can easily be checked that the only nonvanishing component of the conformal curvature tensor is
\begin{gather}
C_{\mu \nu \rho \sigma} l^\mu m^\nu l^\rho m^\sigma
= 2 (b+ic),
\end{gather}
so that we have $b=c=0$. Plugging this to (\ref{b2eqcf}), we obtain
\begin{gather}
a(u) = \pi~\bar{\Phi}(u) \Phi(u), 
\end{gather} 
so that
\begin{gather}
ds^2 = -2\pi \bar \Phi(u) \Phi(u) (x^2 + y^2) du^2 
- 2 du dv  \nn\\
+ dx^2 + dy^2.
\end{gather}
Moreover, the restricted connection and curvature 
tensor are given by
\begin{gather}
\Gm_\mu =-2\pi y~\bar \Phi \Phi~\pd_\mu u,
~~~\tGm_\mu =-2\pi x~\bar \Phi \Phi~\pd_\mu u, \nn\\
K_\mn =2\pi~\bar \Phi \Phi~(\pd_\mu u~\pd_\nu y -\pd_\nu u~\pd_\mu y), \nn\\
\tK_\mn =2\pi~\bar \Phi \Phi~(\pd_\mu u~\pd_\nu x 
-\pd_\nu u~\pd_\mu x).
\end{gather} 
So, with the restricted metric $\cJ_\mn$ and 
$\ctJ_\mn$ given by (\ref{b2rm}) we obtain 
the solution of the $B_2$ gravity which was first discovered by Cahen and Leroy \cite{cahen}.
This is the unique conformally flat solution of 
the Einstein-Maxwell equations which admits 
a covariantly constant null vector field \cite{mclenaghan}.  

\section{Spin-One Graviton}

One of the remarkable features of the restricted gravity is that the metric compatibility condition
is expressed in the form of source-free Maxwell equations. By the Poincare's lemma, a dual 
pair of spin-one fields naturally arise as 
the solutions to these equations. Furthermore, 
as we discussed in the previous section, 
the spacetimes with the trivial topology of $B_2$-gravity belong to the class of pp-wave spacetimes. These facts strongly suggest that 
the spin-one gauge field which appeared as 
the solution to the Maxwellian metric compatibility condition can be interpreted as a representation of 
the graviton that propagates in spacetime. 

For the illustrative purpose, let us examine 
the pp-wave spacetime in the Bondi's coordinates \cite{rosen,bondi},
\begin{gather}
ds^2 = -dt^2 + L^2(e^{2\theta} dx^2 + e^{-2\theta} dy^2)+dz^2,
\label{bondicoord}
\end{gather}
where $L$ and $\theta$ are functions of $u=t-z$. To apply this to the restricted $B_2$-gravity, let us choose the following tetrads
\begin{gather}
e_0 = \pd_t,~~~e_1 = L^{-1} e^{-\theta}~\pd_x, \nn\\ 
e_2 = L^{-1} e^{\theta}~\pd_y,~~~e_3 = \pd_z. 
\label{pwtet}
\end{gather}
With this we have the torsion-free spin connections given by
\begin{gather}
\Gm_\mu= \frac{{\sqrt 2}e^\theta}{L^2} 
\big(L^{\prime} -L \theta^{\prime} \big) \pd_\mu y, \nn\\
~~~\tGm_\mu =\frac{{\sqrt 2} e^{-\theta}}{L^2} 
\big( L^{\prime} +L \theta^{\prime} \big)\pd_\mu x, 
\end{gather}
so that
\begin{gather}
\Gm_\mn = \frac{2\sqrt{2} e^{\theta}}{L^{3}}
\big(LL^{\prime \prime} -L^{2} \theta^{\prime\prime} 
- 2L^{\prime 2}
-L^{2} \theta^{\prime 2} + 2 LL^{\prime} \theta^{\prime} \big) \nn\\
\times (\pro_{[\mu} t~\pro_{\nu]} y -\pro_{[\mu} z~\pro_{\nu]} y).
\nn\\
\tGm_{\mu \nu} = \frac{2\sqrt{2} e^{-\theta}}{L^{3}}
\big(LL^{\prime \prime} + L^{2} \theta^{\prime\prime} - 2L^{\prime 2}
-L^{2} \theta^{\prime 2} - 2 LL^{\prime} \theta^{\prime} \big) \nn\\
\times (\pro_{[\mu} t~\pro_{\nu]} x 
-\pro_{[\mu} z~\pro_{\nu]} x).
\label{fieldstrengthpp}
\end{gather}
Moreover, from (\ref{b2rct}), (\ref{orthobasis}), and (\ref{pwtet}) we have 
\begin{gather}
\cJ_\mn =- {\sqrt{2}} L {e^{-\theta}} 
(\pro_{[\mu} t~\pro_{\nu]} y - \pro_{[\mu} y~\pro_{\nu]} z), \nn\\
\ctJ_\mn  ={\sqrt{2}} L {e^{\theta}} 
(\pro_{[\mu} t~\pro_{\nu]} x+\pro_{[\mu} z~\pro_{\nu]} x),  \nn\\
K_\mu = \Gm_\mu,~~~\tilde{K}_\mu = \tGm_\mu.
\label{restrictedmetricpp}
\end{gather}
so that the first equation of (\ref{b2Req1}) reduces to
\begin{gather}
\Gm_\mn \cJ^{\nu \rho} = \tGm_\mn \ctJ^{\nu \rho}.
\label{pweq1}
\end{gather}
Using (\ref{fieldstrengthpp}), (\ref{restrictedmetricpp}), and (\ref{pweq1}), we obtain
\begin{gather}
L''+L\theta'^2=0.
\label{bondi1}
\end{gather}
This is exactly the gravitational wave equation first obtained by Bondi \cite{bondi}. 

Notice that the Abelian part of the anti-symmetrized four-indexed metric $\cJ^\mn$ and $\ctJ^\mn$ are divergence-free so that there exist a pair of 
the four-vector fields $\cJ_\mu$ and $\ctJ_\mu$ 
such that $\cJ_\mn =2\pd_{[\mu} \cJ_{\nu]}$
and $\ctJ_\mn = 2\pd_{[\mu} \ctJ_{\nu]}$. Explicitly, we have
\begin{gather}
\cJ_\mu =-\sqrt 2 \big(\int L e^{-\theta}du \big) 
~\pd_\mu y, \nn\\
\ctJ_\mu =\sqrt 2 \big(\int L e^{\theta}du \big) 
~\pd_\mu x.
\end{gather}
To understand the physical meaning of these fields, let us consider the linearized version of the pp-wave spacetimes in which the governing equation (\ref{bondi1}) approximates to
\begin{gather}
L^{\prime \prime} \approx 0,
\end{gather}
so that the metric (\ref{bondicoord}) becomes
\begin{gather}
ds^2 = -dt^2 +(1+2\theta) dx^2 +(1-2\theta)dy^2 
+ dz^2.
\end{gather}
This is the plane gravitational wave of polarization $e_+$ propagating along the $z$-direction, known 
as the Rosen-Bondi gravitational wave \cite{rosen,bondi}. The presence of the factor $2$ 
in the $\theta$-terms clearly shows its spin-two character in the deformation of a test-ring on the $xy$-plane. 

In the gauge-theoretical viewpoint, however, we have 
the spin-one fields $\cJ_\mu$ and $\ctJ_\mu$ 
that satisfies the Maxwellian equation. In 
the linearized theory, these fields are
\begin{gather}
\cJ_\mu \approx -\sqrt{2}\big(1 -\theta(u_0) \big) (u-u_0)~\pd_\mu y, \nn\\
\ctJ_\mu \approx \sqrt{2}\big(1 +\theta(u_0) \big) (u-u_0)~\pd_\mu x,
\end{gather}
where $u_0$ is constant. The corresponding components of the anti-symmetrized four-indexed metric along 
the directions of $B_2$ symmetries 
are given by
\begin{gather}
\cJ_\mn \approx -{\sqrt 2} \big(1 -\theta(u_0) \big)
(\pd_{[\mu} t~\pd_{\nu]} y -\pd_{[\mu} y~\pro_{\nu]} z), \nn\\
\ctJ_\mn \approx {\sqrt 2} \big(1 -\theta(u_0) \big)
(\pd_{[\mu} t~\pro_{\nu]} x+\pd_{[\mu} z~\pro_{\nu]} x).
\end{gather}
This confirms that the Rosen-Bondi gravitational wave which represents the graviton at the classical level can be described by a spin-one gauge potential. This has a deep implication in quantum gravity \cite{cqg12,cqg13,gc15}.

One might object this view, because the spin-one gauge field is well known to have a repulsive 
(as well as the attractive) interaction. But as we have emphasized in section II, the gauge theory of Einstein's gravity is not the ordinary 
gauge theory, and thus has different dynamics. 
And we emphasize that here the spin-one field $\cJ_\mu$ (or $\ctJ_\mu$) never generates a repulsive interaction as the gauge potential 
in QED. In the Maxwell theory, the exterior derivative of the Maxwell field which is linear in the Maxwell field itself, determines 
the repulsiveness or the attractiveness of 
the interaction. But in the restricted gravity, however, the equation of motion is different. 
In particular, the acceleration of a test particle is proportional to the restricted 
connection $\hvGm_\mu$, not the spin-one 
field $\cJ_\mu$ or $\ctJ_\mu$. And obviously, $\hvGm_\mu$ shows the attractive nature of 
the interaction since it is identified with 
the Levi-Civita spin-connection of Einstein's theory. 

\section{Discussions}

It has been well known that we can treat Einstein's theory of gravity as the gauge of Lorentz group. Adopting this view and making 
the Abelian decomposition of the gravitational connection, we can decompose Einstein's theory 
to the Abelian part and the valence part, 
and can construct the restricted gravity which describes the core dynamics of Einstein's 
gravity with the Abelian part. There are two types of restricted gravity, the $A_2$ and 
$B_2$ gravity, because the Lorentz group has 
two maximally Abelian subgroups \cite{ijmpa09,cqg12,cqg13,gc15}. 

The Abelian decomposition reveals important hidden structures of Einstein's theory which 
we cannot see in the Einstein's theory. In particular, the Abelian decomposition tells 
that we can have a generally invariant theory of gravity, the restricted gravity, which is simpler than Einstein's theory. However, this was done within the framework of Einstein's theory, with the Abelian decomposition of Einstein's theory.

In this paper we have shown that the restricted gravity can be made a theory of gravity by itself, independent of Einstein's theory, constructing the Lagrangian formalism of 
the restricted gravity. This tells that we can 
view the restricted gravity as an independent theory of gravity, without any reference to Einstein's theory. But one might ask if the restricted gravity can describe a meaningful physics. 

To show that this is so, we have constructed 
important solutions of the restricted gravity using 
the equation of motion of the restricted gravity. 
In particular,we have shown that the $A_2$ gravity 
has the gravitational cosmic string and Berttoti-Robinson solution. Moreover we have shown 
that the $B_2$ gravity has the axisymmetric pp-wave, conformally flat wave with planar wavefront, and 
the Rosen-Bondi gravitational wave. This confirms 
that the restricted gravity can be treated as 
a respectable theory of gravity.

It should be mentioned that we have obtained 
the above solutions of the restricted gravity 
assuming that the restricted gravity has trivial topology. This needs clarification. When we make 
the Abelian decomposition of Einstein's theory, we have to select the Abelian direction of the Lorentz group first. And the Abelian directions in these restricted gravities can have non-trivial topology, which retains the topolgical properties of Einstein's theory in 
the restricted gravity and becomes an essential part of the restricted gravity which plays important roles in the theory. 

But in this paper we have discussed the restricted gravity with the trivial topology for simplicity, 
and have shown that the restricted gravity does have interesting solutions. Of course, one can ask if 
the restricted gravity which has a non-trival topology also has intersting solutions. This is a very important question worth pursuing, and we will discuss this issue in a separate paper \cite{cho}.     

Einstein's theory is described by the spin-two 
metric. Similarly, the restricted gravity is described by the restricted metric. So the restricted metric becomes the fundamental field of the theory. But as 
we have emphasized, the restricted metric can be expressed by a Maxwell type spin-one gauge potential. In other words, the gravity in the restricted gravity
is described by a spin-one gauge potential. This may have a deep implication in quantum gravity, because this opens the possibility that the graviton could be described by a spin-one field. This could provide 
an alternative way of quantizing Einstein's gravity. 

At first thought this view may sound heretic, 
because the gravity is attractive while the spin-one gauge interaction includes a repulsive (as well as attractive) interaction. But as we have pointed out, this is not true. First of all, although the restricted metric is described by a spin-one gauge potential, the equation of motion of the restricted gravity is different from the equation of the gauge theory. So the two theories describe totally different dynamics. Second, the metric is not the only field which describes the graviton. It is well known that in the presence of fermions, the tetrad (i.e., four spin-one fields) describes the gravitational field. This tells that the idea of a spin-one graviton is 
not heretic. Third, it is easier quantizing 
the spin-one field than quantizing the spin-two 
field. This is a big advantage. Moreover, the massless spin-one field has the right degrees of freedom for the massless graviton. Just as the massless spin-two metric, it has two physical degrees. 

Of course, the idea of a spin-one gravition need 
more discussions. Nevertheless, our study in this 
paper strongly implies that this is a very interesting 
idea worth to be discussed further.   

{\bf ACKNOWLEDGEMENT}

~~~The work is supported in part by the National Research Foundation of Korea funded by the Ministry 
of Science and Technology (Grant 2022-R1A2C1006999) 
and by Center for Quantum Spacetime, Sogang University, Korea.

\end{document}	

\bibitem{ijmpa09} Y. M. Cho, Int. J. Mod. Phys. {\bf A24}, 3327 (2009).
\bibitem{cqg12} Y. M. Cho, S. H. Oh, and Sangwoo Kim, Class. Quant. Grav. {\bf 29}, 205007 (2012). 
\bibitem{cqg13} Y. M. Cho, Franklin H. Cho, and J. H. Yoon, 
Class. Quant. Grav. {\bf 30}, 055003 (2013). 
\bibitem{gc15} Y. M. Cho, S. H. Oh, and B. S. Park,
Gravitation and Cosmology {\bf 21}, 257 (2015).

\bibitem{prd76} Y. M. Cho, Phys. Rev. {\bf D14}, 3335 (1976).
\bibitem {hehl} See also, F. Hehl, J. McCrea, E. Mielke,
and Y. Ne'eman, Phys. Rep. {\bf 258}, 1 (1995).

\bibitem{prd80} Y. M. Cho, Phys. Rev. {\bf D21}, 1080 (1980);
Phys. Rev. Lett. {\bf 46}, 302 (1981); Phys. Rev. {\bf D23}, 2415 (1981). See also Y. S. Duan and M. L. Ge, Sci. Sinica {\bf 11},
1072 (1979).
\bibitem{plb14} N. Cundy, Y. M. Cho, W. Lee, and J. Leem, Phys. Lett. {\bf B729}, 192 (2014); Nucl. Phys. {\bf B895}, 64 (2015).
 
\bibitem{prd13} Y. M. Cho, Franklin H. Cho, and J. H. Yoon, 
Phys. Rev. {\bf D87}, 085025 (2013).
\bibitem{epjc19} Y. M. Cho and Franklin H. Cho,  
Euro Phys. J. {\bf C 79}, 498 (2019).

\bibitem{feyn} R. Feynman, Acta. Phys. Pol. {\bf 24}, 697 (1963);
B. DeWitt, Phys. Rev. {\bf 160}, 1113 (1967); {\bf 162}, 1195 (1967). 

\bibitem{fadd} L. Faddeev and A. Niemi, Phys. Rev. Lett.
{\bf 82}, 1624 (1999); Phys. Lett. {\bf B449}, 214 (1999).

\bibitem{shab} S. Shabanov, Phys. Lett. {\bf B458}, 
322 (1999); {\bf B463}, 263 (1999); H. Gies, Phys. Rev. 
{\bf D63}, 125023 (2001).
\bibitem{zucc} R. Zucchini, Int. J. Geom. Meth. Mod. Phys. 
{\bf 1}, 813 (2004).
\bibitem{kondo} K. Kondo, S. Kato, A. Shibata, 
and T. Shinohara, Phys. Rep. {\bf 579}, 1 (2015).

\bibitem{witt} E. Witten, Nucl. Phys. {\bf B249}, 557 (1985). 
\bibitem{bondi} H. Bondi, Nature {\bf 179}, 1072 (1957); 
H. Bondi, F. Pirani, and I. Robinson, Proc. Roy. Soc. (London) 
{\bf A125}, 519 (1959).

\bibitem{cho} Y. M. Cho, Sangwoo Kim, S. H. Oh, to be published.

\bibitem{ehlers} J. Ehlers and W. Kundt, {\it Exact solutions
of the gravitational field equations}, in Gravitation: an 
introduction to current research, ed. L. Witten, Wiley (1962).

\bibitem{nariai} H. Nariai, {\it On some static solutions of 
Einstein's gravitational field equations in a spherically 
symmetric case}, Sci. Rep. Tohoku Univ. 34: 160 (1950).

\bibitem{anariai} H. Nariai, {\it On a new cosmological solution 
of Einstein's field equations of gravitation}, Sci. Rep. Tohoku 
Univ. 35: 62 (1951).

\bibitem{bertotti} B. Bertotti. {\it Uniform Electromagnetic Field 
in the Theory of General Relativity}, Phys. Rev. {\bf 116}, 1331-1333
(1959).

\bibitem{robinson} I. Robinson, {\it A Solution of the Maxwell-Einstein 
Equations}, Bull. Acad. Polon. Sci. {\bf 7}, 351-352 (1959).

\bibitem{hiscock} W. Hiscock, Phys. Rev. {\bf D31}, 3288 (1985).

\bibitem{gott} J. R. Gott III, Astrophys. J. {\bf 288}, 422 (1985).

\bibitem{bonnor} W. B. Bonnor, {\it The gravitational field of light}, 
Comm. Math. Phys. {\bf 13}, 163 (1969).

\bibitem{cahen} M. Cahen and J. Leroy, {\it Exact solutions of 
Einstein-Maxwell equations}, J. Math. Mech. {\bf 16}, 501 (1966).  

\bibitem{roche} C. Roche, A. B. Aazami, and C. Cederbaum, {\it Exact
parallel waves in general relativity}, Gen. Rel. Grav. {\bf 55}, 40
(2023).

\bibitem{mclenaghan} R. G. Mclenaghan, N. Tariq, and B. O. J. Tupper,
{\it Conformally flat solutions of the Einstein-Maxwell equations for
null electromagnetic fields}, J. Math. Phys. {\bf 16}, 829 (1975).
\bibitem{cho} Y. M. Cho, Sangwoo Kim, and S. H. Oh, 
to be published.

\end{thebibliography}
\end{document}